\documentclass[superscriptaddress,notitlepage,amsmath,amssymb]{revtex4-1}
\usepackage{graphicx}

\begin{document}

\title{Schwinger--Dyson equation for quarks in a QCD inspired model}

\author{V. I. Shilin}
\affiliation{Joint Institute for Nuclear Research, Dubna, Russia}
\affiliation{Moscow Institute of Physics and Technology, Dolgoprudny, Russia}
\author{V. N. Pervushin}
\affiliation{Joint Institute for Nuclear Research, Dubna, Russia}

\begin{abstract}
We discuss formulation of QCD in Minkowski--spacetime and effect of an operator product expansion by means of normal ordering of fields in
the QCD Lagrangian. The formulation of QCD in the Minkowski--spacetime allows us to solve a constraint equation and decompose the gauge field propagator in
the sum of an instantaneous part, which forms a bound state, and a retarded part, which contains the relativistic corrections. In Quantum Field Theory, for
%if we not start with Lagrangian as normal ordering function of all operator fields,
a Lagrangian with unordered operator fields, one can make normal ordering by means of the operator product expansion, then
the gluon condensate appear. This gives us a natural way of obtaining a dimensional parameter in QCD, which is missing in the QCD Lagrangian. We
derive a Schwinger--Dyson equation for a quark, which is studied both numerically and analytically. The critical value of the strong coupling
constant $\alpha_s = 4/{\pi}$, above which a nontrivial solution appears and a spontaneous chiral symmetry breaking occurs, is found. For the
sake of simplicity, the considered model describes only one flavor massless quark, but the methods can be used in more general
case. The Fourier-sine transform of a function with log-power asymptotic was performed.
\end{abstract}

\maketitle

\section{Introduction.}
Strong interaction physics should be described by Quantum Field Theory (QFT) with the Quantum Chromodynamic (QCD)
Lagrangian \cite{YangMills54,FaddeevPopov67,Faddeev69,tHooft71_RanormYangMills,GrossWilczek73_AssymptFreedom,Politzer73_AssymptFreedom}. As it
shown \cite{GrossWilczek73_AssymptFreedom,Politzer73_AssymptFreedom}, the running coupling constant $\alpha_s$ is strong enough at small
energies, so that perturbation expansion is not applicable. This is a big problem due to the lack of general methods of non-perturbative calculations.

In order to describe the strong interaction, the phenomenological models were developed that are based directly on experimental data and use partly
the QCD knowledge: the QCD sum rules \cite{ShifmanVainshteinZakharov79_1_QCDSumRules,ShifmanVainshteinZakharov79_2_QCDSumRules,Ioffe81_QCDSumRules}, the Chiral
Perturbation Theory \cite{VolkovPervushin79_book,VolkovPervushin75_ChPTPEPAN,VolkovPervushin76_ChPTUFN,GasserLeutwyler_ChPT}, the Nambu--Jona-Lasinio model and its
generalizations \cite{NambuJonaLasinio61_NJL1,VolkovEbert83,Volkov84,Volkov86_review,Reinhardt88_NJLVtH,Bernard88_NJLVtH2,Yaouanc85_NJLVnonlocal}, bag
models \cite{BogolubovPN68_bag,Chodos74_bag1,Chodos74_bag2} and others \cite{Haymaker91,Pawlowski07_FuncRenormGroup}. These
models can relatively easily reproduce experimental data. However, they have a number of disadvantages: each of these models works in a
certain application area but fail in others, the accuracy of  theoretical calculations are limited and often less than the accuracy of modern experimental
data. And these models are not true theory of strong interactions. This gives
impetus to construct models based directly on QCD, for instance: instanton liquid
model \cite{Shuryak82_Instanton1,Shuryak82_Instanton2,Shuryak82_Instanton3,Dyakonov84_Instanton}, domain wall
network \cite{Nedelko95_DomainWall,Nedelko96_DomainWall,Nedelko01_DomainWall,Nedelko04_DomainWall,Nedelko11_DomainWall,Nedelko15_DomainWall_Impact}, various estimations from
Schwinger--Dyson equations \cite{Roberts94_SDeq_QED3_QCD,Alkofer97_SDeq,Alkofer01_SDeq_QED3_QCD,Roberts03_SDeq_rev,Fischer06_SDeq_rev, Reinhardt07_SDeq_Hamiltonian,Aguilar08_SDeq}.

Thus, there exist various approximations to the theory of strong interactions with their specific simplifications of the QCD. In the QCD researches, one
should find answers to the key questions, which are the description of QCD vacuum, spontaneous breaking of chiral symmetry, the absence
of color particles (confinement problem), the description of bound states, their masses and decay widths.

We consider the theory of strong interaction at low energy. Our aim is to emphasize the importance of formulation in Minkowski-spacetime and effect of
an operator product expansion by means of normal ordering of fields in Lagrangian, and to discuss some consequences of this novel approach.

The formulation of QCD in the Minkowski-spacetime allows us to solve a constraint equation and decompose the gauge field propagator in the sum of an
instantaneous part, which forms a bound state, and a retarded part, which contains the relativistic corrections. At the first stage, we should neglect the retarded part
and use the instantaneous part to construct the bound state. Then the retarded part gives corrections to the already existing bound state. This idea cames
form QED \cite{ShilinPervushin13_QED} (see also \cite{Polubarinov03,Pervushin03_DAAD,Pervushin90_proc}), where any attempts of working with the entire propagator
do not lead to satisfactory results or the decomposition occurs implicitly. Our approach enable us to cover the both high- and low-energy ranges and find the relation
between fundamental QCD parameters and low energy constants.

In QFT, for a Lagrangian with unordered operator fields, one can make normal ordering by means of the operator product expansion. Then
the gluon condensate and a low energy effective gluon mass appear. This mechanism gives us a natural and fundamental way of obtaining a dimensional parameter in QCD, which
is missing in the QCD Lagrangian. The existence of non-zero condensates directly linked to the conformal anomaly of QCD.

In the next section, we start from QCD Lagrangian and derive an effective action of strong interaction. Then
in section \ref{section_SD_equation}, by using this effective action we obtain the Schwinger--Dyson equation for a quark, which is solved both
numerically (in section \ref{section_numerical_solution}) and analytically (in the subsequent sections). In conclusion, we summarize the obtained
results and discuss the prospects of the developed methods. Here, for
the sake of simplicity, we intentionally neglect some effects, for example, the considered model describes only one flavor massless quark. While
investigating of the Schwinger--Dyson equation, we focus mainly on the question of the spontaneous symmetry breaking. Nevertheless, all
the assumptions made to derive the equation are transparent and well-controlled.

\section{Effective Action for the Strong Interaction.\label{section_effective_action}}

Let us start with the Quantum Chromodynamics Lagrangian, with number of colors $N_c = 3$ and number of flavors $N_f = 1$:
\begin{equation}
\mathcal{L}_\mathrm{QCD}= -\frac{1}{4}F^a_{\mu\nu} F^{a\mu\nu} - A^a_{\mu} j^{a\mu} + \overline{\psi}( i\gamma^\mu \partial_\mu - m ) \psi \; ,
\label{lagrangian_QCD_initial}
\end{equation}
where $A^a_{\mu}$, $F^a_{\mu\nu} = \partial_{\mu}A^a_{\nu} - \partial_{\nu}A^a_{\mu} + g f^{abc} A^b_\mu A^c_\nu$, $\psi$, $m$,
and $j^{a\mu} = -g \overline{\psi} \gamma^\mu \frac{\lambda^a}{2} \psi$ are the gluon field, gluon field strength tensor, quark field, quark current
mass, and color current of quark, respectively.

An effective action for the meson-like bound state can be derived
from the Lagrangian (\ref{lagrangian_QCD_initial}). To this end, some \emph{restrictions} and \emph{assumptions} are needed. Below the
symbol \labelitemi \ is introduced for convenience when we discuss another one assumption or resrtiction. Some of the restrictions are not principle but imposed
in order to not overload the reader by technical calculations. Anyway, in the developed model, we outline main
ideas that may be important for correct description of meson-like bound state rather than give a complete description of strong
interaction, which certainly remains a tremendous problem.

\begin{itemize}
\item First, we choose the frame of reference where the bound state, which we obtain and discuss below, is as whole at rest. Therefore
only the \emph{static} problems are considered. We emphasize that the proper choice of the reference frame should be done in Minkowski-spacetime rather than
in Euclidian-spacetime. Note that the generalization of this theory to one bound state moving on mass
shell \cite{Pervushin03_DAAD, PervushinShilin13_QCD, Pervushin90_proc,Pervushin12_proc,Pervushin15_proc} can easily be done, it is
sufficient to rewrite various quantities in the comoving frame of reference.
\end{itemize}

We fix the gauge
\begin{equation}
\partial_k A^a_k(x) =0 \: ,
\label{gluon_gauge}
\end{equation}
where $k=1,2,3$ and $a = 1, \ldots, 8$ are the space and gluon color indexes, respectively.

The gluon term in the Lagrangian takes the form
\begin{multline}
-\frac{1}{4}F^a_{\mu\nu} F^{a\mu\nu} = \\
= \frac{1}{2}\dot{A}^a_i\dot{A}^a_i - \frac{1}{4}F^a_{ij}F^{aij} + gf^{abc} (\partial_0A^a_i) A^b_0 A^c_i -
\frac{1}{2} A^a_0 \, \partial_i \partial_i A^a_0 - gf^{abc} (\partial_iA^a_0) A^b_0 A^c_i + \frac{1}{2} g^2 f^{abc} f^{ade} A^b_0 A^c_i A^d_0 A^e_i \; .
\label{gluon_lagr_before_no}
\end{multline}
The third term on the right hand side contains the time derivative and thus can be neglected, because only the static problems are considered, as is noted above.

After quantization, the gluon field $A^a_\mu$ becomes an operator field. One can consider the vacuum 2-point correlator
\begin{equation}
\langle 0| A^a_i(x) A^b_j(x) |0 \rangle = 2C_g \delta_{ij} \delta^{ab} \: .
\label{gloun_correlator}
\end{equation}
\begin{itemize}
\item We assume that $C_g \neq 0$ and $C_g < \infty$. Actually $C_g$ depends on the energy, but for simplicity we suppose $C_g$ to be a constant. The
constant $C_g$ can be determined from a phenomenology.
\end{itemize}
The fields in Eq.~(\ref{gloun_correlator}) obeys  the condition (\ref{gluon_gauge}). A question at once arises: ``How should the
formula (\ref{gloun_correlator}) be rewritten in other gauges?'' The answer is to make a gauge transform to (\ref{gluon_gauge}), then impose the
condition (\ref{gloun_correlator}), and then make the inverse gauge transform. As is said above, when solving the on-shell bound state problem, we always
have a privileged frame of reference, in which this bound state as whole is at rest; therefore, we always have the
privileged gauge (\ref{gluon_gauge}), and thus we define (\ref{gloun_correlator}) in a gauge-covariant manner in this way. Note that physically privileged
reference frame is absent for any scattering problem of quarks and gluons, and one cannot define (\ref{gloun_correlator}) the same way.

Usually in Quantum Field Theory, a Lagrangian contains only normally ordered operator fields. This is a result of normal ordering of an initial Lagrangian where
the above correlator-like terms, arising due to the ordering, are omitted, because they are considered as (infinite) vacuum energy contributions. Keeping
these terms, we have after the normal ordering
\begin{equation}
-\frac{1}{4}F^a_{\mu\nu} F^{a\mu\nu} =\;\,
:\!\frac{1}{2}\dot{A}^a_i\dot{A}^a_i\!: \,-\, :\!\frac{1}{4}F^a_{ij}F^{aij}\!: \,+\, :\!\frac{1}{2} A^a_0 (-\Delta + M_g^2) A^a_0\!: \,+\; \ldots \; .
\label{gluon_lagr_after_no}
\end{equation}
The term with $M_g$ comes from the last term of formula (\ref{gluon_lagr_before_no}) and  $M_g^2 \equiv 6 g^2 C_g N_c$. Here we use
the relation $f^{acb} f^{acd} = N_c \delta^{bd}$. The quantity $M_g$ might be interpreted as an effective gluon mass in the gauge (\ref{gluon_gauge}). This
is essentially a model-dependent quantity. In this approach, the gluon mass appears before a perturbation
expansion. Phenomenological models in which gluons have nonzero effective mass at small energies have been considered earlier by some
authors (see \cite{Aguilar08_SDeq,Mandula87_GluonMassLattice,Amemiya99_GluonMassLattice,Cornwall83_GluonMassGlueball,Pervushin90_GluonMass,
Halzen93_GluonMassPomeron,Larin13_GluonMass,Larin15_GluonMass,Larin16_GluonMass} and references therein).
\begin{itemize}
\item Let us consider the dotted terms in (\ref{gluon_lagr_after_no}) as a perturbation and neglect them. This assumption means that we suggest that
bound states are \emph{formed} by only some of the terms which explicitly written in expression (\ref{gluon_lagr_after_no}), while the other
terms merely give some \emph{corrections} to the already existing bound states. In the basic model developed in this paper, these terms are neglected. The
neglected terms can influence on quantitative characteristics of the bound states, but not their presence, and numerical amount of
corrections might be not small due to large value of strong coupling constant.
\end{itemize}

Substituting (\ref{gluon_lagr_after_no}) without dotted terms into  Lagrangian (\ref{lagrangian_QCD_initial}), we arrive at the generating functional
\begin{multline*}
\displaystyle \mbox{\large $\displaystyle \mathcal{Z} = \int \mathsf{D}A^a_\mu \delta(\partial_k A^a_k) \mathsf{D}\overline{\psi} {\mathsf D}\psi \exp $} \biggl[
i \int d^4\!x \Bigl( \frac{1}{2}\dot{A}^a_i\dot{A}^a_i - \frac{1}{4}F^a_{ij}F^{aij} + \frac{1}{2} A^a_0 (-\Delta + M_g^2) A^a_0 - A^a_0 j^a_0 + A^a_i j^a_i + \\
+ \overline{\psi}( i\gamma^\mu \partial_\mu - m ) \psi \Bigr) + i \int d^4\!x ( A^a_i J^{ai} + \overline{\eta}\psi + \overline{\psi}\eta) \biggr] \; .
\end{multline*}
The source $J^a_0$ is not involved, since the field $A^a_0$ is not dynamical degree of freedom with the gauge (\ref{gluon_gauge}). This is owing to the fact that
the corresponding equation of motion is a constraint \cite{Dirac55}.

Making integration over $A^a_0$ yields
\begin{multline*}
\displaystyle \mbox{\large $\displaystyle \mathcal{Z} = \int \mathsf{D}A^a_k \delta(\partial_k A^a_k) \mathsf{D}\overline{\psi} {\mathsf D}\psi \exp $} \biggl[
i \int d^4\!x \Bigl( \frac{1}{2}\dot{A}^a_i\dot{A}^a_i - \frac{1}{4}F^a_{ij}F^{aij} + A^a_i j^a_i + \overline{\psi}( i\gamma^\mu \partial_\mu - m ) \psi \Bigr) - \\
- \frac{i}{2} \int d^4\!x \, d^4\!y \; j^a_0(x) \delta(x^0 \! - \! y^0) \frac{1}{4\pi}
\frac{e^{- M_g |\mathbf{x} - \mathbf{y}|}}{|\mathbf{x} \! - \! \mathbf{y}|} j^a_0(y) + i \int d^4\!x ( A^a_i J^{ai} + \overline{\eta}\psi + \overline{\psi}\eta) \biggr] \; .
\end{multline*}
The term
$$
- \frac{1}{2} \int d^4\!x \, d^4\!y \; j^a_0(x) \delta(x^0 \! - \! y^0) \frac{1}{4\pi}
\frac{e^{- M_g |\mathbf{x} - \mathbf{y}|}}{|\mathbf{x} \! - \! \mathbf{y}|} j^a_0(y)
$$
includes a combination of the Gell-Mann matrices, which may be rewritten in the form
$$
\frac{\lambda^{a r_1 r_2}}{2} \frac{\lambda^{a s_2 s_1}}{2} = \frac{1}{3} \delta^{r_1 s_1} \delta^{r_2 s_2} + \frac{1}{6} \varepsilon^{t r_1 s_2} \varepsilon^{t s_1 r_2} \; .
$$
\begin{itemize}
\item We restrict ourselves to the colorless mesons and so neglect the second term. This term is the diquark channel, which plays a role when baryons are taken into
account ("baryon = diquark + quark").
\end{itemize}
Thus within this approximation, the above term can be rewritten in the form
\begin{multline*}
-\frac{1}{2} \int d^4\!x_1 \, d^4\!x_2 \, d^3\!\mathbf{y}_1 \, d^3\!\mathbf{y}_2 \;
\overline{\psi}^{r_1}_{\alpha_1}(x_1) \, \psi^{\alpha_2 r_2}(x_2) \, \delta^{r_1 s_1} \times \\
\times \underbrace{{\gamma^0}^{\alpha_1}_{\phantom{\alpha_1}\alpha_2} \, \delta^4\!(x_1 \! - \! x_2) \,
\frac{g^2}{12\pi}\frac{e^{-M_g |\mathbf{x}_1 - \mathbf{y}_2|}}{|\mathbf{x}_1 \! - \! \mathbf{y}_2|} \,
\delta^3\!(\mathbf{y}_1 \! - \! \mathbf{y}_2) \, {\gamma^0}^{\beta_2}_{\phantom{\beta_2}\beta_1}
}_{\mathcal{K}^{\alpha_1 \phantom{\beta_1 \alpha_2} \beta_2}_{\phantom{\alpha_1} \beta_1 \alpha_2}(x_1, \mathbf{y}_1 ; x_2, \mathbf{y}_2)}
\delta^{r_2 s_2} \, \overline{\psi}^{s_2}_{\beta_2}(x^0_2,\mathbf{y}_2) \, \psi^{\beta_1 s_1}(x^0_1, \mathbf{y}_1) =
\end{multline*}
$$
= -\frac{1}{2} \int d^4\!x_1 \, d^3\!\mathbf{y}_1 \, d^4\!x_2 \, d^3\!\mathbf{y}_2 \;
\overline{\psi}^r_{\alpha_1}(x_1) \, \psi^{\beta_1 r}(x^0_1, \mathbf{y}_1) \:
\mathcal{K}^{\alpha_1 \phantom{\beta_1 \alpha_2} \beta_2}_{\phantom{\alpha_1} \beta_1 \alpha_2}(x_1, \mathbf{y}_1 ; x_2, \mathbf{y}_2) \:
\psi^{\alpha_2 s}(x_2) \, \overline{\psi}^s_{\beta_2}(x^0_2,\mathbf{y}_2) \; .
$$
where the above formula is the definition of the
operator $\mathcal{K}^{\alpha_1 \phantom{\beta_1 \alpha_2} \beta_2}_{\phantom{\alpha_1} \beta_1 \alpha_2}(x_1, \mathbf{y}_1 ; x_2, \mathbf{y}_2)$,
and $\psi^{\beta_1 s_1}(x^0_1, \mathbf{y}_1)$ was shifted to the left. One can see that color indexes $r$ and $s$ have been summed inside
pairs $\psi\overline{\psi}$, so the pair $\psi\overline{\psi}$ as whole is colorless.
\begin{itemize}
\item Let us treat $\psi^{\alpha s}(x^0,\mathbf{x}) \, \overline{\psi}^s_{\beta}(x^0, \mathbf{y})$ as a real bilocal field.
\end{itemize}
The operator $\mathcal{K}^{\alpha_1 \phantom{\beta_1 \alpha_2} \beta_2}_{\phantom{\alpha_1} \beta_1 \alpha_2}(x_1, \mathbf{y}_1 ; x_2, \mathbf{y}_2)$ is symmetrical
and has an inverse operator $\mathcal{K}^{-1}$ that can be defined by:
$$
\int d^4\!x_2 \, d^3\!\mathbf{y}_2 \;
\mathcal{K}^{\alpha_1 \phantom{\beta_1 \alpha_2} \beta_2}_{\phantom{\alpha_1} \beta_1 \alpha_2}(x_1, \mathbf{y}_1 ; x_2, \mathbf{y}_2)
\: {\mathcal{K}^{-1}}^{\alpha_2 \phantom{\beta_2 \alpha_3} \beta_3}_{\phantom{\alpha_2} \beta_2 \alpha_3}
(x_2, \mathbf{y}_2 ; x_3, \mathbf{y}_3) \; = \;
\delta^4\!(x_1 \! - \! x_3) \, \delta^3\!(\mathbf{y}_1 \! - \! \mathbf{y}_3) \,
\delta^{\alpha_1}_{\phantom{\alpha_1}\alpha_3} \, \delta^{\beta_3}_{\phantom{\beta_3}\beta_1} \; .
$$
This allows us to introduce new bilocal field ${\mathcal M}^{\alpha}_{\phantom{\alpha} \beta}(x^0, \mathbf{x}, \mathbf{y})$ and make a bosonization
transform (Habbard-Stratanovich transform) \cite{PervushinEbert76_conf2,PervushinEbert76_conf,Kleinert76_BilocalHadronization,
PervushinReinhardtEbert79,Stratonovich57_HSTransform,Hubbard59_HSTransform}:
\begin{multline*}
\mbox{\large $\displaystyle \exp $} \biggl[-\frac{i}{2} \int d^4\!x_1 \, d^3\!\mathbf{y}_1 \, d^4\!x_2 \, d^3\!\mathbf{y}_2 \;
\overline{\psi}^r_{\alpha_1}(x_1) \, \psi^{\beta_1 r}(x^0_1, \mathbf{y}_1) \:
\mathcal{K}^{\alpha_1 \phantom{\beta_1 \alpha_2} \beta_2}_{\phantom{\alpha_1} \beta_1 \alpha_2}(x_1, \mathbf{y}_1 ; x_2, \mathbf{y}_2) \:
\psi^{\alpha_2 s}(x_2) \, \overline{\psi}^s_{\beta_2}(x^0_2,\mathbf{y}_2) \biggr] \mbox{\large $\displaystyle = $} \\[\jot]
\shoveleft{\mbox{\large $\displaystyle = \int \mathsf{D}\mathcal{M} \exp $} \biggl[
\frac{i}{2} \int d^4\!x_1 \, d^3\!\mathbf{y}_1 \, d^4\!x_2 \, d^3\!\mathbf{y}_2 \;
\mathcal{M}^{\mathrm T}{}_{\alpha_1}^{\phantom{\alpha_1} \beta_1}(x^0_1, \mathbf{x}_1, \mathbf{y}_1) \:
{\mathcal{K}^{-1}}^{\alpha_1 \phantom{\beta_1 \alpha_2} \beta_2}_{\phantom{\alpha_1} \beta_1 \alpha_2}(x_1, \mathbf{y}_1 ; x_2, \mathbf{y}_2)
\: \mathcal{M}^{\alpha_2}_{\phantom{\alpha_2} \beta_2}(x^0_2, \mathbf{x}_2, \mathbf{y}_2) \;+ }\\
+\; i \int d^4\!x \, d^3\!\mathbf{y} \; \overline{\psi}^r_{\alpha}(x^0, \mathbf{x}) \, \psi^{\beta r}(x^0,\mathbf{y}) \:
\mathcal{M}^{\alpha}_{\phantom{\alpha} \beta}(x^0, \mathbf{x}, \mathbf{y}) \biggr] \; .
\end{multline*}

Finally the generating functional for effective action of strong interaction takes the form
\begin{multline}
\displaystyle \mbox{\large $\displaystyle \mathcal{Z} = \int \mathsf{D}A^a_k \delta(\partial_k A^a_k) \mathsf{D}\overline{\psi} {\mathsf D}\psi \mathsf{D}\mathcal{M} \exp $}
\biggl[ i \int d^4\!x \Bigl( \frac{1}{2}\dot{A}^a_i\dot{A}^a_i - \frac{1}{4}F^a_{ij}F^{aij} + A^a_i j^a_i + \overline{\psi}( i\gamma^\mu \partial_\mu - m ) \psi \Bigr) + \\
+ \frac{i}{2} \int d^4\!x_1 \, d^3\!\mathbf{y}_1 \, d^4\!x_2 \, d^3\!\mathbf{y}_2 \;
\mathcal{M}^{\mathrm T}{}_{\alpha_1}^{\phantom{\alpha_1} \beta_1}(x^0_1, \mathbf{x}_1, \mathbf{y}_1) \:
{\mathcal{K}^{-1}}^{\alpha_1 \phantom{\beta_1 \alpha_2} \beta_2}_{\phantom{\alpha_1} \beta_1 \alpha_2}(x_1, \mathbf{y}_1 ; x_2, \mathbf{y}_2)
\: \mathcal{M}^{\alpha_2}_{\phantom{\alpha_2} \beta_2}(x^0_2, \mathbf{x}_2, \mathbf{y}_2) \; + \\
+\; i \int d^4\!x \, d^3\!\mathbf{y} \; \overline{\psi}^r_{\alpha}(x^0, \mathbf{x}) \, \psi^{\beta r}(x^0,\mathbf{y}) \:
\mathcal{M}^{\alpha}_{\phantom{\alpha} \beta}(x^0, \mathbf{x}, \mathbf{y})
+ i \int d^4\!x ( A^a_i J^{ai} + \overline{\eta}\psi + \overline{\psi}\eta) \biggr] \; .
\label{generating_functional_final}
\end{multline}
With the help of this generating functional, one can write down any diagrams for processes of interest.

\section{Schwinger--Dyson equation.\label{section_SD_equation}}

In what follows we restrict ourselves only to the question of spontaneous symmetry breaking in the theory described by the
functional (\ref{generating_functional_final}). For this purpose, it is convenient to derive and investigate the Schwinger--Dyson (Gap) equation for the quark.

It is difficult to examine the Schwinger--Dyson equation in the general form.
\begin{itemize}
\item For the sake of simplicity, we use the Stationary Phase method (that is the Semiclassical approximation). This
method simplify the Schwinger--Dyson equation but retain its main properties.
\end{itemize}
According to this method, we should integrate out the Fermion variables $\psi$ and $\overline{\psi}$ in (\ref{generating_functional_final}), thus  deriving
the functional for the action $S_{eff}$
\begin{equation}
\mathcal{Z} = \int \mathsf{D}A^a_k \delta(\partial_k A^a_k) \mathsf{D}\mathcal{M} \, e^{i S_{eff}} \; .
\label{generating_functional_integrated}
\end{equation}
Then we arrive at the Schwinger--Dyson equation
\begin{equation}
\frac{\delta S_{eff}}{\delta {\mathcal M}} (A^a_k=0, \overline{\eta}=0, \eta=0, J=0) = 0 \; ,
\label{SDequation_definition}
\end{equation}
which gives us the Fermion spectrum inside the bound
state \cite{PervushinReinhardtEbert79,PervushinEbert76_conf2,PervushinEbert76_conf,PervushinReinhardtEbert79,KKPervushinSarikov90_F,KKPervushin89_Y,KKPervushin90_F,KKPervushin89_P,
Pervushin03_DAAD}.

We introduce the operator
$$
{G_{mA\mathcal{M}}^{-1}}^{\alpha\phantom{\beta}rs}_{\phantom{\alpha}\beta}(x,y)\equiv
\left( i {\gamma^\mu}^{\alpha}_{\phantom{\alpha} \beta} \delta^{rs} \partial_\mu - m\delta^{\alpha}_{\phantom{\alpha} \beta}\delta^{rs}
+gA^a_i {\gamma^i}^{\alpha}_{\phantom{\alpha} \beta}\frac{\lambda^{ars}}{2}\right) \delta^4(x\!-\!y) \;+\;
{\mathcal M}^{\alpha}_{\phantom{\alpha} \beta}(x^0, \mathbf{x}, \mathbf{y}) \, \delta(x^0\!-\!y^0) \, \delta^{rs}
$$
and define its inverse as
$$
\int d^4\!y \; {G_{mA\mathcal{M}}^{-1}}^{\alpha\phantom{\beta}rs}_{\phantom{\alpha} \beta}(x,y) \:
{G_{mA\mathcal{M}}}^{\beta\phantom{\gamma}st}_{\phantom{\beta} \gamma}(y,z) \;=\;
\delta^{\alpha}_{\phantom{\alpha} \gamma} \, \delta^{rt} \, \delta^4(x-z) \; .
$$
In this notations, formula (\ref{generating_functional_integrated}) reads
\begin{multline*}
\displaystyle \mbox{\large $\displaystyle \mathcal{Z} = \int \mathsf{D}A^a_k \delta(\partial_k A^a_k) \mathsf{D}\mathcal{M} \exp $}
\biggl[ i \int d^4\!x \Bigl( \frac{1}{2}\dot{A}^a_i\dot{A}^a_i - \frac{1}{4}F^a_{ij}F^{aij} \Bigr) + \\
+ \frac{i}{2} \int d^4\!x_1 \, d^3\!\mathbf{y}_1 \, d^4\!x_2 \, d^3\!\mathbf{y}_2 \;
\mathcal{M}^{\mathrm T}{}_{\alpha_1}^{\phantom{\alpha_1} \beta_1}(x^0_1, \mathbf{x}_1, \mathbf{y}_1) \:
{\mathcal{K}^{-1}}^{\alpha_1 \phantom{\beta_1 \alpha_2} \beta_2}_{\phantom{\alpha_1} \beta_1 \alpha_2}(x_1, \mathbf{y}_1 ; x_2, \mathbf{y}_2)
\: \mathcal{M}^{\alpha_2}_{\phantom{\alpha_2} \beta_2}(x^0_2, \mathbf{x}_2, \mathbf{y}_2) \; - \\
-\; i \int d^4\!x \, d^4\!y \; \overline{\eta}(x) \: G_{mA\mathcal{M}}(x,y) \: \eta(y) \; + \; \mathrm{tr}\ln G_{mA\mathcal{M}}^{-1}
\; + \; i \int d^4\!x A^a_i J^{ai} \biggr] \; .
\end{multline*}
Inserting the corresponding $S_{eff}$ into equation (\ref{SDequation_definition}) we arrive at
\begin{equation}
\int d^4\!x_2 \, d^3\!\mathbf{y}_2 \;
{\mathcal{K}^{-1}}^{\alpha_1 \phantom{\beta_1 \alpha_2} \beta_2}_{\phantom{\alpha_1} \beta_1 \alpha_2}(x_1, \mathbf{y}_1 ; x_2, \mathbf{y}_2)
\: \mathcal{M}^{\alpha_2}_{\phantom{\alpha_2} \beta_2}(x^0_2, \mathbf{x}_2, \mathbf{y}_2) \;+\;
i \int dy^0_1 \; {G_{mA\mathcal{M}}}^{\alpha_1\phantom{\beta_1}r}_{\phantom{\alpha_1} \beta_1\phantom{r}r}(x_1, y_1) \Bigr|_{A=0} \! \delta(x^0_1\!-\!y^0_1) = 0 \; .
\label{SDequation_1step}
\end{equation}
Below in this article, the solution of this equation is denoted by
$$
\mathcal{M}^{\alpha}_{\phantom{\alpha} \beta}(x^0, \mathbf{x}, \mathbf{y}) \;=\;
- \Sigma^{\alpha}_{\phantom{\alpha} \beta}(x^0, \mathbf{x}, \mathbf{y}) \;+\;
m \delta^{\alpha}_{\phantom{\alpha} \beta} \, \delta^3(\mathbf{x}\!-\!\mathbf{y}) \; .
$$
It is convenient to introduce the operator
$$
{G_{\Sigma}^{-1}}^{\alpha}_{\phantom{\alpha} \beta}(x,y) \equiv
i{\gamma^\mu}^{\alpha}_{\phantom{\alpha} \beta} \partial_\mu \,\delta^4(x\!-\!y) \;-\;
\Sigma^{\alpha}_{\phantom{\alpha} \beta}(x^0, \mathbf{x}, \mathbf{y}) \, \delta(x^0\!-\!y^0) \; ,
$$
which, on the stationary solutions obeying Eq.~(\ref{SDequation_1step}), coincides with the earlier introduced operator
$${G_{mA\mathcal{M}}^{-1}}^{\alpha\phantom{\beta}rs}_{\phantom{\alpha}\beta}(x,y)\Bigr|_{A=0}\;=\;{G_{\Sigma}^{-1}}^{\alpha}_{\phantom{\alpha}\beta}(x,y)\delta^{rs}.$$
The inverse operator is defined in the standard manner
$$
\int d^4\!y \; {G_{\Sigma}^{-1}}^{\alpha}_{\phantom{\alpha} \beta}(x,y) \,
{G_{\Sigma}}^{\beta}_{\phantom{\beta} \gamma}(y,z) \;=\;
\delta^{\alpha}_{\phantom{\alpha} \gamma}\delta^4(x-z) \; .
$$
Acting with the operator $\mathcal{K}$ on the both sides of Eq.~(\ref{SDequation_1step}) and using the above notations, we obtain
\begin{equation}
\Sigma^{\alpha_1}_{\phantom{\alpha_1} \beta_1}(x^0_1, \mathbf{x}_1, \mathbf{y}_1) \;=\;
m \, \delta^{\alpha_1}_{\phantom{\alpha_1} \beta_1} \, \delta^3(\mathbf{x}_1\!-\!\mathbf{y}_1) \;+\;
3i \int d^4\!x_2 \, d^4\!y_2 \;
\mathcal{K}^{\alpha_1 \phantom{\beta_1 \alpha_2} \beta_2}_{\phantom{\alpha_1} \beta_1 \alpha_2}(x_1, \mathbf{y}_1 ; x_2, \mathbf{y}_2)
\: {G_{\Sigma}}^{\alpha_2}_{\phantom{\alpha_2} \beta_2}(x_2, y_2) \, \delta(x^0_2\!-\!y^0_2) \; .
\label{SDequation_2step}
\end{equation}

\begin{itemize}
\item We are looking for a simplest solution of this equation and adopt the following ansatz
$$
\Sigma^{\alpha}_{\phantom{\alpha} \beta}(x^0, \mathbf{x}, \mathbf{y}) =
\delta^\alpha_{\phantom{\alpha} \beta} \frac{1}{(2\pi)^\frac{3}{2}} M(\mathbf{x}\!-\!\mathbf{y}) \; .
$$
Due to the isotropy,  $M$ is radially symmetric and depends only on $|\mathbf{x}\!-\!\mathbf{y}|$.
\end{itemize}

Making the Fourier transform of equation (\ref{SDequation_2step}), we have
\begin{equation}
M(\mathbf{p}) \, \delta^{\alpha_1}_{\phantom{\alpha_1} \beta_1} \;=\;
m \, \delta^{\alpha_1}_{\phantom{\alpha_1} \beta_1} \;-\;
i \frac{g^2}{(2\pi)^4} \int d^4\!q \; \frac{1}{(\mathbf{p}-\mathbf{q})^2 + M_g^2}
\: \gamma^{0 \alpha_1}_{\phantom{0 \alpha_1} \alpha_2} \, {G_{\Sigma}}^{\alpha_2}_{\phantom{\alpha_2} \beta_2}(q) \, \gamma^{0 \beta_2}_{\phantom{0 \beta_2} \beta_1} \; .
\label{SDequation_3step}
\end{equation}
In momentum space, the operator $G_{\Sigma}^{-1}$ can easily be reversed
$$
G_{\Sigma}(q) = e^{-\gamma^i \frac{q_i}{|\mathbf{q}|} \varphi(\mathbf{q})} \biggl(\frac{1}{q_0 + E(\mathbf{q}) - i\varepsilon} \cdot \frac{1 + \gamma^0}{2} \;+\;
\frac{1}{q_0 - E(\mathbf{q}) + i\varepsilon} \cdot \frac{1 - \gamma^0}{2}\biggr) e^{\gamma^i \frac{q_i}{|\mathbf{q}|} \varphi(\mathbf{q})} \gamma^0 \; ,
$$
where we put by definition  $E(\mathbf{q}) \equiv \sqrt{{M(\mathbf{q})}^2 + \mathbf{q}^2}$ , and: $\cos 2\varphi(\mathbf{q}) \equiv \frac{M(\mathbf{q})}{E(\mathbf{q})}$. One
can see that the time-component $q_0$ appears in Eq.~(\ref{SDequation_3step}) only through $G_{\Sigma}(q)$ and, hence, can be integrated out.
$$
M(\mathbf{p}) = m +
\frac{\pi g^2}{(2\pi)^4} \int d^3\!\mathbf{q} \frac{1}{(\mathbf{p}-\mathbf{q})^2 + M_g^2} \frac{M(\mathbf{q})}{E(\mathbf{q})} \; .
$$
After integrating over the solid angle in 3D momentum space, finally the Schwinger--Dyson equation takes the form
\begin{equation}
M(p) = m + \frac{g^2}{(4\pi)^2}\frac{1}{p} \int\limits_{0}^{\infty} dq \frac{q M(q)}{\sqrt{M^2(q) + q^2}} \ln \biggl( \frac{M_g^2 + (p+q)^2}{M_g^2 + (p-q)^2} \biggr) \; ,
\label{SDequation_final_with_m}
\end{equation}
with $p \equiv |\mathbf{p}|$ and $q \equiv |\mathbf{q}|$ being the absolute values of $\mathbf{p}$ and $\mathbf{q}$, respectively.

As discussed above, Eq.~(\ref{SDequation_definition}) describes a fermion spectrum inside the bound state. Thus, the physical meaning of $M(p)$ is a running
quark mass, and, hence, it should be positive for any momentum. At $p=0$, the value $M(0)$ corresponds to a constituent quark mass, while the current quark
mass is $m$. One can introduce instead of quark charge $g$ a strong coupling constant $\alpha_s \equiv {g^2}/{(4\pi)}$. It
is well known in QCD $\alpha_s$ is a running coupling, whose value strongly dependent of the energy scale. Moreover, at low energies, the dependence
of momentum $\alpha_s (p)$ can not be calculated from the perturbation theory, which is inapplicable due to the large value of $\alpha_s$. In the literature, there
exist various predictions about the shape of $\alpha_s (p)$ (see, e.g.,
\cite{Pawlowski07_FuncRenormGroup,Alkofer97_SDeq,Alkofer01_SDeq_QED3_QCD,Roberts03_SDeq_rev,Fischer06_SDeq_rev,Reinhardt07_SDeq_Hamiltonian,Ilgenfritz09_LatticeRunCoupling} and
references therein). Nevertheless in this paper, we assume that $\alpha_s$ is a constant; which is consistent with, as we mention above, neglecting corrections to the
bound states; this means in particular neglecting all the loop corrections to $\alpha_s$, and $\alpha_s$ is really a constant in the framework of this approach. So
in a way, the used in this article constant $\alpha_s$ can be understood as an average of the strong coupling $\alpha_s(p)$ over $p$ within a low-momentum range.

\begin{itemize}
\item We solve the equation (\ref{SDequation_final_with_m}) only for $m=0$, which can be justified by the phenomenology. Indeed, $m \ll M(0)$, because the current mass
of light $u$ and $d$ quarks is about $5 \, \mathrm{MeV}$. On the other hand, the constituent mass of the same quarks is of order $300 \, \mathrm{MeV}$ for different models.
\item One can demand $M(q) \to 0$ when $q \to \infty$. Due to the asymptotic freedom at large momenta, the running quark mass tends to the current mass. Although the
existence of the asymptotic freedom in our model is questionable, we do not want to violate it explicitly. In addition, if this restriction is fulfilled then
the equation (\ref{SDequation_final_with_m}) does \emph{not} need any renormalization.
\end{itemize}

It is convenient to introduce the dimensionless variables $\bar{p}\equiv{p}/{M_g}$, $\bar{q}\equiv{q}/{M_g}$, and $\bar{M}(\bar{p})\equiv{M(p)}/{M_g}$. In this
variables, the Schwinger--Dyson equation (\ref{SDequation_final_with_m}) takes the form
\begin{equation}
\bar{M}(\bar{p}) = \frac{g^2}{(4\pi)^2}\frac{1}{\bar{p}} \int\limits_{0}^{\infty} d\bar{q} \frac{\bar{q} \bar{M}(\bar{q})}{\sqrt{\bar{M}^2(\bar{q})+\bar{q}^2}}
\ln\biggl(\frac{1+(\bar{p}+\bar{q})^2}{1+(\bar{p}-\bar{q})^2}\biggr) \; .
\label{SDequation_final}
\end{equation}
It is obvious that there always exists the solution $\bar{M}(\bar{p}) = 0$. Of course, we are looking for a nontrivial solution of this equation.

An attractive feature of the Schwinger--Dyson equation (\ref{SDequation_final}) is that it is controlled only by one external parameter $g$, which should be fixed
from the phenomenology. In particular, it follows from the definition of the dimensionless variables and Eq.~(\ref{SDequation_final}) that the
constituent quark mass is a linear function of $M_g$ and the coefficient of proportionality $c(g)$ depends only on $g$: $M(0) = c(g) \cdot M_g$.

\section{Numerical solution of the Schwinger--Dyson equation.\label{section_numerical_solution}}

To solve equation (\ref{SDequation_final}) numerically, we use the following algorithm. Let us take a zeroth-order approximation function $\bar{M}_0(\bar{p})$, it
is desirable that $\bar{M}_0(\bar{p})$ differs from the solution $\bar{M}(\bar{p})$ not much. Then substituting $\bar{M}_0(\bar{p})$ into the integral in
the right-hand side (\ref{SDequation_final}) we get $\bar{M}_1(\bar{p})$ in the left-hand side. Then $\bar{M}_1(\bar{p})$ is substituted again, and so on. After a certain number
of steps we get, up to the errors of computer calculations, the exact solution $\bar{M}(\bar{p})$ for which the substitution into the right-hand side (\ref{SDequation_final}) gives
itself. Strictly speaking, we should prove that this algorithm is convergent. We did not try to prove this because in all cases that we calculated
this algorithm turned out to be convergent. Moreover, there is no difference in the choice of $\bar{M}_0(\bar{p})$ (see below for details).

In some sense, the convergence of the algorithm can be explained by the stability of the solution under small perturbations. Namely, substituting
the function $(1+\varepsilon) \bar{M}(\bar{p})$, where $\varepsilon \ll 1$, into integral (\ref{SDequation_final}) we have up to $\varepsilon^2$ terms
\begin{multline*}
\frac{g^2}{(4\pi)^2}\frac{1}{\bar{p}} \int\limits_{0}^{\infty} d\bar{q} \frac{\bar{q} (1+\varepsilon)\bar{M}(\bar{q})}{\sqrt{(1+\varepsilon)^2\bar{M}^2(\bar{q})+\bar{q}^2}}
\ln\biggl(\frac{1+(\bar{p}+\bar{q})^2}{1+(\bar{p}-\bar{q})^2}\biggr) \simeq \\
\simeq (1+\varepsilon) \bar{M}(\bar{p}) -
\varepsilon\frac{g^2}{(4\pi)^2}\frac{1}{\bar{p}} \int\limits_{0}^{\infty} d\bar{q} \frac{\bar{q} \bar{M}^3(\bar{q})}{(\sqrt{\bar{M}^2(\bar{q})+\bar{q}^2})^3}
\ln\biggl(\frac{1+(\bar{p}+\bar{q})^2}{1+(\bar{p}-\bar{q})^2}\biggr) \; .
\end{multline*}
The last term is smaller than $\varepsilon \bar{M}(\bar{p})$ and has a minus sign. That is why the obtained expression
is closer to the solution $\bar{M}(\bar{p})$ than $(1+\varepsilon) \bar{M}(\bar{p})$.

After some attempts to solve equation (\ref{SDequation_final}) numerically, we have found that there are
some things that \emph{should be avoided} at numerical computation:
\begin{enumerate}
\item The upper limit of integration must be $+\infty$ and cannot be replaced by finite quantity $\Lambda$; otherwise a strong dependence of the solution form $\Lambda$ appears.
\item $\bar{M}(+\infty)=0$, otherwise the integral diverges.
\item It is better to avoid replacing the continuous function $\bar{M}(\bar{p})$ by a discrete table $\bar{M}(\bar{p}_i)$ with fixed numbers of points $\bar{p}_i$. That is because
the value of $\bar{M}$ at the penultimate point (at the last point $\bar{M}=0$, as it is noted above) depends mainly on the behavior of $\bar{M}(\bar{p})$ between
this point and the end point and has a weak dependence on the values of $\bar{M}$ a the other points; the value of $\bar{M}$ at the next to penultimate
point depends on the value of $\bar{M}$ at the penultimate point and the behavior of $\bar{M}(\bar{p})$ between these three points, and so on. One cannot
approximate the behavior of the function $\bar{M}(\bar{p})$ between two points by a linear segment, otherwise this leads to very low accuracy of numerical
calculations. Preferably, $\bar{M}(\bar{p})$ expands in a series of known functions. One may also add that maybe we have more accurate results than
in paper \cite{Puzynin99}, where a similar equation was considered numerically and such
replacing $\bar{M}(\bar{p})$ by the table $\bar{M}(\bar{p}_i)$ was done.
\end{enumerate}

Put by definition $\bar{M}(-\bar{p}) = \bar{M}(\bar{p})$, then equation (\ref{SDequation_final}) can be rewritten in the form
\begin{equation}
\bar{M}(\bar{p}) = \frac{g^2}{2(4\pi)^2}\frac{1}{\bar{p}} \int\limits_{-\infty}^{+\infty} d\bar{q} \frac{\bar{q} \bar{M}(\bar{q})}{\sqrt{\bar{M}^2(\bar{q})+\bar{q}^2}}
\ln\biggl(\frac{1+(\bar{p}+\bar{q})^2}{1+(\bar{p}-\bar{q})^2}\biggr) \; .
\label{SDequation_whole_axis}
\end{equation}

Let us define the new function
\begin{equation}
W(\bar{q}) \equiv \frac{\bar{q} \bar{M}(\bar{q})}{\sqrt{\bar{M}^2(\bar{q})+\bar{q}^2}} \; .
\label{def_w}
\end{equation}
One can easily see that $W(\bar{q})$ has the properties
\begin{eqnarray}
\bar{q} \,\to\, +\infty: &\qquad& W(\bar{q}) \simeq \bar{M}(\bar{q}) \label{condition_w_1} \; , \\ [\jot]
\bar{q}>0: && 0 \leqslant W(\bar{q}) \leqslant \min{(\bar{q},\bar{M}(\bar{q}))} \label{condition_w_2} \; , \\ [\jot]
&& W(-\bar{q}) = -W(\bar{q}) \; . \nonumber
\end{eqnarray}

New variables can be introduced (where $\lambda$ -- is some parameter)
\begin{eqnarray*}
\displaystyle\bar{p} = \lambda \tan \Bigl(\frac{\varphi}{2}\Bigr) &\quad , \quad& \varphi \in (-\pi,\pi) \; , \\ [\jot]
\displaystyle\bar{q} = \lambda \tan \Bigl(\frac{\theta}{2}\Bigr) &,& \theta \in (-\pi,\pi) \; .
\end{eqnarray*}
In this variables the Schwinger-Dyson equation takes form
\begin{equation}
\bar{M}(\varphi) = \frac{g^2}{2(4\pi)^2} \int\limits_{-\pi}^{+\pi} \frac{d\theta}{2 \, \tan\frac{\varphi}{2} \: \cos^2 \frac{\theta}{2}}
\ln\biggl(\frac{1 + \lambda^2 (\tan\frac{\varphi}{2} + \tan\frac{\theta}{2})^2}{1+ \lambda^2 (\tan\frac{\varphi}{2} - \tan\frac{\theta}{2})^2}\biggr) W(\theta) \; .
\label{SDequation_finite_variables}
\end{equation}

On $[-\pi,\pi]$ there is a convenient system of the Fourier series functions:
\begin{align*}
& \left\{ \begin{aligned} \;
\bar{M}(\varphi) &= \frac{a_0}{2} \: + \: \sum_{k=1}^\infty a_k \cdot \cos (k\varphi) \phantom{\int\limits_{-\pi}} \\
W(\theta) &= \sum_{k=1}^\infty b_k \cdot \sin (k\theta) \phantom{\int\limits^{+\pi}}
\end{aligned} \right. &
\left\{ \begin{aligned} \;\;
a_0 &= \frac{1}{\pi} \int\limits_{-\pi}^{+\pi} \bar{M}(\varphi) \, d\varphi \\
a_k &= \frac{1}{\pi} \int\limits_{-\pi}^{+\pi} \bar{M}(\varphi) \cos (k\varphi) \, d\varphi \\
b_k &= \frac{1}{\pi} \int\limits_{-\pi}^{+\pi} W(\theta) \sin (k\theta) \, d\theta \; .
\end{aligned} \right.
\end{align*}
Using the Fourier series expansion we can avoid all the numerical difficulties which were discussed above. As the Fourier harmonics are periodical functions, it
would be better if the area of the fastest change of the function lay closer to the center of the interval. The point $\theta = \frac{\pi}{2}$ corresponds to $\bar{q} = \lambda$, so
it dictates the choice of $\lambda$. Of course, before the calculation we do not know what value should be taken; fortunately, the incorrect $\lambda$ leads only to
hight inaccuracy and low speed of calculation. Equation (\ref{SDequation_finite_variables}) now takes the matrix form
\begin{equation}
a_k \,=\, A_{kj} b_j \, ,
\label{SDequation_matrix}
\end{equation}
where $\displaystyle A_{kj} \equiv \frac{g^2}{32 \pi^3} M_{kj}$, where
$$
M_{kj} \equiv \int\limits_{-\pi}^{+\pi} d\varphi \int\limits_{-\pi}^{+\pi} d\theta
\frac{\cos(k\varphi)}{2 \, \tan\frac{\varphi}{2} \: \cos^2 \frac{\theta}{2}}
\ln\biggl(1 + \frac{\lambda^2 \sin\varphi \, \sin\theta}{(\cos\frac{\varphi}{2} \, \cos\frac{\theta}{2})^2 + (\lambda \sin\frac{\varphi-\theta}{2})^2}\biggr) \sin(j\theta) \; .
$$
The matrix $M_{kj}$ contains only the parameter $\lambda$ and can be calculated separately. Of course
in computations, the infinite Fourier series are truncated to finite ones with some number $N_h$ of harmonics.

The zeroth-order approximation function $W_0(\varphi)$ should obey at least condition (\ref{condition_w_2}). We tried various $W_0(\varphi)$, which
obey (\ref{condition_w_2}), and in all cases got the same results differing only in evaluation time. That is why one can take $W_0(\bar{q}) = \bar{q}$.

Finally, the algorithm is the following.
$$
\begin{array}{ccccc}
\text{Take } W_0(\varphi) &&&& \\
\downarrow &&&&\\ [\smallskipamount]
\begin{array}{c} \text{Expand } W_0(\varphi) \\ \text{to Fourier coeff. } b_j \end{array} &&&& \\ [\smallskipamount]
\Big\downarrow &&&& \\ [\medskipamount]
\begin{array}{c} \text{Insert in the Schwinger-Dyson} \\ \text{equation (\ref{SDequation_matrix}):   } a_k \,=\, A_{kj} b_j \end{array} & \longrightarrow &
\begin{array}{c} \text{Insert } a_k \text{ in the function:} \\ \bar{M}(\varphi) = \frac{a_0}{2} +\sum_{k=1}^\infty a_k \cdot \cos (k\varphi) \end{array} && \\ [\bigskipamount]
\big\uparrow && \big\downarrow & \searrow & \\ [\medskipamount]
\begin{array}{c} \text{Expand } W(\varphi) \\ \text{ to Fourier coeff. } b_j \end{array} & \longleftarrow &
\begin{array}{c} \text{Insert } \bar{M}(\varphi) \text{ in the function:} \\
W(\varphi) = \frac{\lambda \sin\frac{\varphi}{2} \bar{M}(\varphi)}{\sqrt{(\cos\frac{\varphi}{2} \bar{M}(\varphi))^2 + (\lambda \sin\frac{\varphi}{2})^2}} \end{array} &&
\text{    Result} \\ [\bigskipamount]
\big\downarrow &&& \nearrow & \\ [\medskipamount]
\begin{array}{c} \text{Insert } b_k \text{ in the function:} \\ W(\theta) = \sum_{k=1}^\infty b_k \cdot \sin (k\theta) \end{array}
& \longrightarrow & \text{Insert } W(\bar{q}) \text{ in exact SD equation (\ref{SDequation_final})}
\end{array}
$$

This procedure gives the solution that is expanded into the Fourier series. We can check how well enough it is by substitution in the
exact (not matrix (\ref{SDequation_matrix})) equation (\ref{SDequation_final}).

\begin{figure*}[!htb]
\begin{minipage}{240pt}
\includegraphics[scale=1]{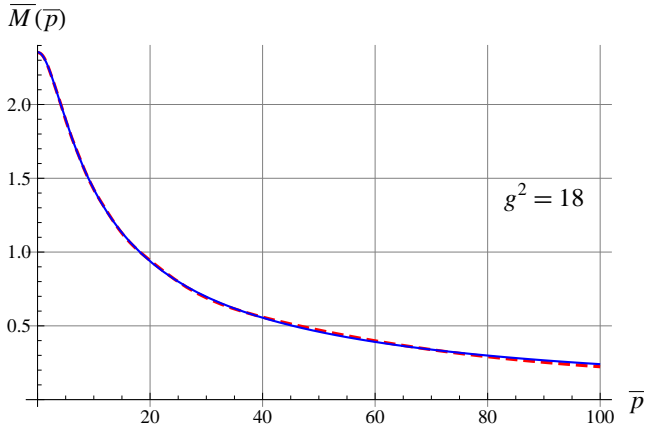}
\end{minipage}
\hfill
\begin{minipage}{240pt}
\includegraphics[scale=1]{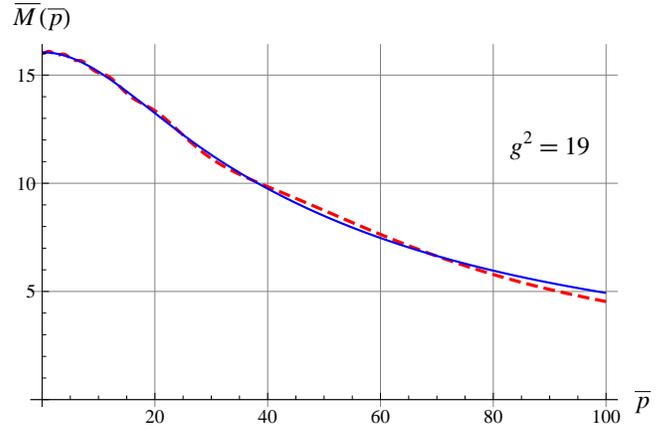}
\end{minipage}
\caption{The running quark mass versus momentum $\bar{M}(\bar{p})$, in units of $M_g$, at different values of the coupling constant $g^2$. The computational parameters
are $\lambda = 10$, $N_h = 13$ (see the explanations in the text). The numerical solution of matrix equation (\ref{SDequation_matrix}) is shown in red dashed thick line. The
blue solid thin line represents the result of substitution of the previous solution into the right-hand side of Schwinger--Dyson equation (\ref{SDequation_final}).}
\label{plot_solution_lb10}
\end{figure*}

The result of numerical research is the following. There is only the trivial solution $\bar{M}(\bar{p}) = 0$ when $g^2 < 16$. The number $16$ is exact and can be obtained from
analytical estimations (see Section \ref{section_analytical_restrictions}). At $g^2 > 16$ a nonzero solution appears. Examples of such calculation are shown
in Fig.~\ref{plot_solution_lb10}. One can see that the above-mentioned check is successful.
%As the result of numerical simulation does not depends on the zeroth-order approximation function, one can say that, there is uniqueness of this nontrivial solution.

\begin{figure*}[!htb]
\begin{minipage}{240pt}
\includegraphics[scale=1]{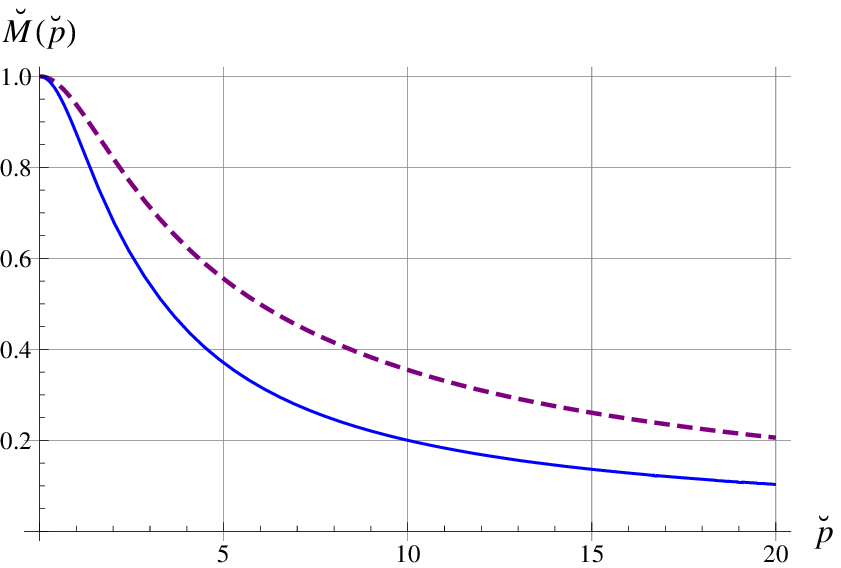}
\caption{The running quark mass versus momentum $\breve{M}(\breve{p})$, in units of constituent quark mass $M(0)$. Actually, plots are the same as in Fig.~\ref{plot_solution_lb10} but
in the breve variables. All lines represent the results of substitutions of the solutions of matrix
equation (\ref{SDequation_matrix}) into the right-hand side of Schwinger--Dyson equation (\ref{SDequation_final}). The purple dashed thick line and the blue solid thin
line correspond to the cases $g^2 = 18$ and $g^2 = 19$ respectively.}
\label{plot_solution_Mbreve_lb10_val13_20}
\end{minipage}
\hfill
\begin{minipage}{240pt}
\includegraphics[scale=1]{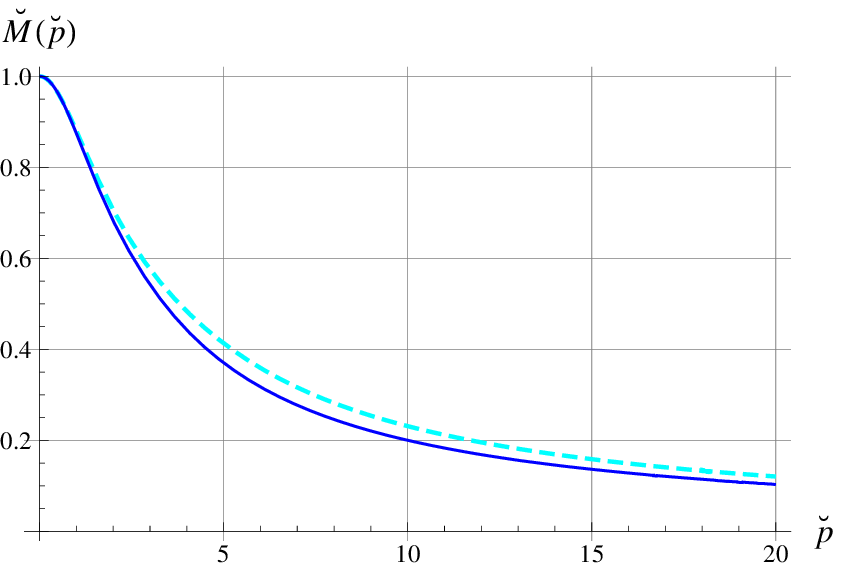}
\caption{The running quark mass $\breve{M}(\breve{p})$ calculated from nonlinear (\ref{SDequation_whole_axis}) and linear (\ref{SDequation_whole_axis_linear}) equations. The
computational parameters are $\lambda = 10$, $N_h = 13$ and $g^2 = 19$. Blue solid thin line is the same as in Fig.~\ref{plot_solution_Mbreve_lb10_val13_20}. The cyan dashed
thick line is the analog for the linear equation.}
\label{plot_solution_MbreveNLandL_lb10_val13_g19_20}
\end{minipage}
\end{figure*}

Unfortunately, due to a low accuracy of the numerical calculations, we can not obtain the precise value $\bar{M}(0)$. Namely, equation (\ref{SDequation_final}) in
the other dimensionless variables $\breve{p} \equiv {p}/{M(0)}$, $\breve{q} \equiv {q}/{M(0)}$ and $\breve{M}(\breve{p}) \equiv {M(p)}/{M(0)}$, takes the form
$$
\breve{M}(\breve{p}) = \frac{g^2}{(4\pi)^2}\frac{1}{\breve{p}} \int\limits_{0}^{\infty} d\breve{q}
\frac{\breve{q} \breve{M}(\breve{q})}{\sqrt{{\breve{M}}^2(\breve{q})+{\breve{q}}^2}} \ln\biggl(1+\frac{4\breve{p}\breve{q}}{\breve{M}_g^2+(\breve{p}-\breve{q})^2}\biggr) \; ,
$$
where $\breve{M}_g \equiv {M_g}/{M(0)} = 1/{\bar{M}(0)}$, and there is the condition $\breve{M}(0) = 1$. One can see that right-hand side of this
equation depends on $\bar{M}(0)$ only logarithmically, and when $|\breve{q}-\breve{p}| \gg \breve{M}_g$ it does not depends on $\bar{M}(0)$ at all. The
solutions of the above equation are shown in Fig.~\ref{plot_solution_Mbreve_lb10_val13_20} for $g^2 - 16 = 2$ and $g^2 - 16 = 3$.

\section{Analytical restrictions.\label{section_analytical_restrictions}}
Using the expression $\bar{M}(\bar{p}) \;=\; -\!\int\limits_{\bar{p}}^{+\infty}\!\! d\tilde{q} \, \bar{M}^\prime(\tilde{q}) \;=\;
-\!\int\limits_{0}^{+\infty}\!\! d\bar{q} \, \bar{M}^\prime(\bar{q}+\bar{p})$ one can rewrite equation (\ref{SDequation_final}) in the form
$$
\int\limits_{0}^{+\infty} d\bar{q} \, \Biggl( \frac{g^2}{(4\pi)^2}\frac{1}{\bar{p}} \frac{\bar{q} \bar{M}(\bar{q})}{\sqrt{\bar{M}^2(\bar{q})+\bar{q}^2}}
\ln\biggl(\frac{1+(\bar{p}+\bar{q})^2}{1+(\bar{p}-\bar{q})^2}\biggr) \:+\: \bar{M}^\prime(\bar{q}+\bar{p}) \Biggr) \;=\; 0 \; .
$$
Upon integrating this equation over $\bar{p}$ from zero to infinity, exchange of the order of integration, and direct integration over $\bar{p}$ we have
\begin{equation}
\int\limits_{0}^{+\infty} d\bar{q} \, \biggl( \frac{\bar{q} \bar{M}(\bar{q})}{\sqrt{\bar{M}^2(\bar{q})+\bar{q}^2}} \frac{g^2}{(4\pi)^2}
2\pi\arctan(\bar{q}) \:-\: \bar{M}(\bar{q}) \biggr) \;=\; 0 \; .
\label{SDequation_integrated}
\end{equation}
The formula is correct even if the integral $\int\limits_{0}^{+\infty} \!\! d\bar{q} \; \bar{M}(\bar{q})$ diverges. That is because in all steps the right-hand side
of the equation is zero.

If $g^2 \leqslant 16$, then: $\frac{g^2}{(4\pi)^2} 2\pi\arctan(\bar{q}) < \frac{g^2}{4^2} \leqslant 1$. So we get
$$
0 \;=\; \int\limits_{0}^{+\infty} d\bar{q} \, \biggl( \frac{\bar{q}}{\sqrt{\bar{M}^2(\bar{q})+\bar{q}^2}} \frac{g^2}{(4\pi)^2}
2\pi\arctan(\bar{q}) \,-\, 1 \biggr) \, \bar{M}(\bar{q}) \;<\;
\int\limits_{0}^{+\infty} d\bar{q} \underbrace{\phantom{\Bigg|}\biggl( \frac{\bar{q}}{\sqrt{\bar{M}^2(\bar{q})+\bar{q}^2}} \,-\, 1 \biggr)}_{< 0}
\cdot \underbrace{\bar{M}(\bar{q})\phantom{\Bigg|}}_{\geqslant 0} \; .
$$
Which can be satisfied only when $\bar{M}(\bar{p}) = 0$ for any $\bar{p}$.

For $g^2 > 16$ we have in the limit $\bar{q} \to \infty$:
$$
\biggl( \frac{\bar{q}}{\sqrt{\bar{M}^2(\bar{q})+\bar{q}^2}} \frac{g^2}{(4\pi)^2} 2\pi\arctan(\bar{q}) \,-\, 1 \biggr) \, \bar{M}(\bar{q}) \;\to\;
\Bigl(\frac{g^2}{16} - 1 \Bigr) \bar{M}(\bar{q}) \; .
$$
This means that $\int\limits_{0}^{+\infty} \!\! d\bar{q} \; \bar{M}(\bar{q}) < \infty$. For $g^2 > 16$, the integrand in (\ref{SDequation_integrated}) is
below zero at small $\bar{q}$ and above zero at large $\bar{q}$, and so the whole integral (\ref{SDequation_integrated}) can be equal to zero.

Thus, for $g^2 \leqslant 16$, we have only the trivial solution $M(p) = 0$ of the
massless Schwinger-Dyson equation (\ref{SDequation_final}). For $g^2 > 16$, the integral $\int\limits_{0}^{+\infty} \!\! d\bar{q} \; \bar{M}(\bar{q})$ is
convergent. The threshold value $g^2 = 16$ corresponds to $\alpha_s = \frac{4}{\pi} \simeq 1.27$. Also
recall that we have taken $N_f = 1$, for other values of $N_f$ the critical value of $\alpha_s$ might be different. It is worthwhile to notice that this critical value
lies near the maximum value $\alpha_s \simeq 1.2$ of the function $\alpha_s (p)$ obtained from the lattice calculations \cite{Ilgenfritz09_LatticeRunCoupling}.

\section{The linearized Schwinger--Dyson equation.}

\subsection{The linearized Schwinger--Dyson equation and its numerical solution.}

Let us introduce the notation $\bar{M}_0 \equiv \bar{M}(0)$ and the function
\begin{equation}
\mathcal{W}(\bar{q}) \equiv \frac{\bar{q} \bar{M}(\bar{q})}{\sqrt{\bar{M}^2_0+\bar{q}^2}} \; .
\label{def_calw}
\end{equation}
Then: $\bar{M}(\bar{p}) = \frac{1}{\bar{p}} \sqrt{\bar{M}^2_0+\bar{p}^2} \, \mathcal{W}(\bar{p})$. The function $\mathcal{W}(\bar{q})$ has the properties
\begin{eqnarray}
\bar{q} \,\to\, \infty: &\qquad& \mathcal{W}(\bar{q}) \simeq W(\bar{q}) \; , \label{condition_mcw_1} \\ [\jot]
\bar{q} \,\to\, 0: && \mathcal{W}(\bar{q}) \simeq W(\bar{q}) \; , \nonumber \\ [\jot]
\bar{q}>0: && 0 \leqslant \mathcal{W}(\bar{q}) < \bar{M}(\bar{q}) \; , \nonumber \\ [\jot]
&& \mathcal{W}(-\bar{q}) = -\mathcal{W}(\bar{q}) \label{condition_mcw_3} \; .
\end{eqnarray}
\begin{itemize}
\item These properties show that $\mathcal{W}(\bar{p})$ can be determined from an approximate to (\ref{SDequation_final}) equation
\begin{equation}
\sqrt{\bar{M}^2_0+\bar{p}^2} \, \mathcal{W}(\bar{p}) \: = \:
\frac{g^2}{(4\pi)^2} \int\limits_0^{+\infty} d\bar{q} \: \ln\biggl(\frac{1+(\bar{p}+\bar{q})^2}{1+(\bar{p}-\bar{q})^2}\biggr) \, \mathcal{W}(\bar{q}) \; .
\label{SDequation_final_linear}
\end{equation}
\end{itemize}

As well as equation (\ref{SDequation_final}) was rewritten in the form (\ref{SDequation_whole_axis}), one can use (\ref{condition_mcw_3}) and
then equation (\ref{SDequation_final_linear}) takes form
\begin{equation}
\sqrt{\bar{M}^2_0+\bar{p}^2} \, \mathcal{W}(\bar{p}) \: = \:
\frac{g^2}{2(4\pi)^2} \int\limits_{-\infty}^{+\infty} d\bar{q} \: \ln\biggl(\frac{1+(\bar{p}+\bar{q})^2}{1+(\bar{p}-\bar{q})^2}\biggr) \, \mathcal{W}(\bar{q}) \; .
\label{SDequation_whole_axis_linear}
\end{equation}
The last equation can be solved numerically in the same way as in section \ref{section_numerical_solution} equation (\ref{SDequation_whole_axis}) was
solved, only one need to use (\ref{def_calw}) instead of (\ref{def_w}). The result of such calculation is shown
in Fig.~\ref{plot_solution_MbreveNLandL_lb10_val13_g19_20} together with the numerical solution of the nonlinear equation (\ref{SDequation_whole_axis}). The
plots of the solutions of the nonlinear and linear equations turn out to be close.

\subsection{Analytical restrictions for the linearized Schwinger--Dyson equation.}

Equation (\ref{SDequation_final_linear}) is the eigenvalue problem, where $g^2$ plays role the inverse eigenvalue. There is the Perron--Frobenius theorem
for a real square matrix with positive entries. It might be supposed that analog of this theorem holds true for equation (\ref{SDequation_final_linear}). Then we can
conclude, equation (\ref{SDequation_final_linear}) has the unique strictly positive eigenvector, and the corresponding $g^2$ is real positive smallest inverse eigenvalue.

Let us assume that at fixed $g^2 > 16$, equation (\ref{SDequation_final_linear}) has two different nontrivial
solutions $\mathcal{W}(\bar{p})$ and $\mathcal{W}^\prime(\bar{p})$ for different $\bar{M}_0$ and $\bar{M}_0^\prime$ (let be $\bar{M}_0 > \bar{M}_0^\prime$), respectively,
\begin{equation}
\left\{ \begin{aligned} \;
\sqrt{\bar{M}^2_0+\bar{p}^2} \, \mathcal{W}(\bar{p}) \: & = \:
\frac{g^2}{(4\pi)^2} \int\limits_0^{+\infty} d\bar{q} \: \ln\biggl(\frac{1+(\bar{p}+\bar{q})^2}{1+(\bar{p}-\bar{q})^2}\biggr) \, \mathcal{W}(\bar{q}) \; , \\
\sqrt{\bar{M}^{\prime 2}_0+\bar{q}^2} \, \mathcal{W}^\prime(\bar{q}) \: & = \:
\frac{g^2}{(4\pi)^2} \int\limits_0^{+\infty} d\bar{p} \: \ln\biggl(\frac{1+(\bar{p}+\bar{q})^2}{1+(\bar{p}-\bar{q})^2}\biggr) \, \mathcal{W}^\prime(\bar{p}) \; .
\label{SDequation_whole_axis_linear_2solutions}
\end{aligned} \right.
\end{equation}
Upon multiplying the first equation by the function $\mathcal{W}^\prime(\bar{p})$, integrating the result over $\bar{p}$ from zero to plus infinity and
exchange of the order of integration we have
$$
\int\limits_0^{+\infty} d\bar{p} \: \mathcal{W}^\prime(\bar{p}) \, \sqrt{\bar{M}^2_0+\bar{p}^2} \, \mathcal{W}(\bar{p}) \: = \:
\frac{g^2}{(4\pi)^2} \int\limits_0^{+\infty} d\bar{q} \int\limits_0^{+\infty} d\bar{p} \:
\mathcal{W}(\bar{q}) \ln\biggl(\frac{1+(\bar{p}+\bar{q})^2}{1+(\bar{p}-\bar{q})^2}\biggr) \, \mathcal{W}^\prime(\bar{p}) \; .
$$
Making integration over $\bar{p}$ by substituting the second equation of the set (\ref{SDequation_whole_axis_linear_2solutions}) we arrive at
$$
\int\limits_0^{+\infty} d\bar{p} \, \underbrace{ \phantom{\Big|} \!\! \mathcal{W}^\prime(\bar{p})}_{> 0}
\underbrace{ \Bigl( \sqrt{\bar{M}^2_0+\bar{p}^2} \, - \, \sqrt{\bar{M}^{\prime 2}_0+\bar{p}^2} \Bigr) }_{> 0}
\underbrace{ \mathcal{W}(\bar{p}) \phantom{\Big|} \!\! }_{> 0} \, = \: 0 \; .
$$
The last formula is the contradiction. Consequently $\bar{M}_0 = \bar{M}_0^\prime$,
in other words there is the single $\bar{M}_0$ for which equation (\ref{SDequation_final_linear}) admits the positive solution. And
vice versa, for any $g^2 > 16$ there is the single $\bar{M}_0 > 0$ for which the unique positive solution exist.

\subsection{The linearized Schwinger--Dyson equation in the Fourier space.}

In equation (\ref{SDequation_whole_axis_linear}) in the numerator of the logarithm one can change the
variable $\bar{q} \mapsto -\bar{q}$. The Schwinger--Dyson equation (\ref{SDequation_whole_axis_linear}) then takes the form
\begin{equation}
\sqrt{\bar{M}^2_0+\bar{p}^2} \, \mathcal{W}(\bar{p}) \: = \:
-\frac{g^2}{(4\pi)^2} \int\limits_{-\infty}^{+\infty} d\bar{q} \: \ln\Bigl(1+(\bar{p}-\bar{q})^2\Bigr) \, \mathcal{W}(\bar{q}) \; .
\label{SDequation_mcw_p}
\end{equation}

The right-hand side of the last equation can be simplified by means of the Fourier transform
\begin{align*}
\mathcal{W}(x) \, &= \; \frac{1}{\sqrt{2\pi}} \int\limits_{-\infty}^{+\infty} d\bar{p} \: e^{i\bar{p}x} \, \mathcal{W}(\bar{p}) \; , &
\mathcal{W}(\bar{p}) \, &= \; \frac{1}{\sqrt{2\pi}} \int\limits_{-\infty}^{+\infty} dx \: e^{-i\bar{p}x} \, \mathcal{W}(x) \; ,
\end{align*}
\begin{multline*}
\frac{1}{\sqrt{2\pi}} \int\limits_{-\infty}^{+\infty} d\bar{p} \: e^{i\bar{p}x}
\int\limits_{-\infty}^{+\infty} d\bar{q} \: \ln\Bigl(1+(\bar{p}-\bar{q})^2\Bigr) \, \mathcal{W}(\bar{q}) \:= \\
=\; \frac{1}{\sqrt{2\pi}} \int\limits_{-\infty}^{+\infty} d\bar{q} \int\limits_{-\infty}^{+\infty} d\bar{p} \: e^{i\bar{q}x} e^{i(\bar{p}-\bar{q})x} \,
\mathcal{W}(\bar{q}) \ln\Bigl(1+(\bar{p}-\bar{q})^2\Bigr) \:= \\
=\; \sqrt{2\pi} \, \biggl( \frac{1}{\sqrt{2\pi}} \int\limits_{-\infty}^{+\infty} d\bar{q} \: e^{i\bar{q}x} \, \mathcal{W}(\bar{q}) \biggr) \cdot
\biggl( \frac{1}{\sqrt{2\pi}} \int\limits_{-\infty}^{+\infty} d\bar{p} \: e^{i\bar{p}x} \ln\Bigl(1+\bar{p}^2\Bigr) \biggr) \:=\;
- \, 2\pi \, \frac{e^{-|x|}}{|x|} \, \mathcal{W}(x) \; .
\end{multline*}
It should be noted that since $\mathcal{W}(\bar{p})$ is an odd function (\ref{condition_mcw_3}), the Fourier transform reduces
to a Fourier-sine transform $\mathcal{W}(x) = i \mathcal{W}_s(x)$, where:
\begin{align}
\mathcal{W}_s(x) \, &= \; \sqrt{\frac{2}{\pi}} \int\limits_{0}^{+\infty} d\bar{p} \: \sin(\bar{p}x) \, \mathcal{W}(\bar{p}) &
\mathcal{W}(\bar{p}) \, &= \; \sqrt{\frac{2}{\pi}} \int\limits_{0}^{+\infty} dx \: \sin(\bar{p}x) \, \mathcal{W}_s(x) \; .
\label{transform_sin-Fourier}
\end{align}
The Schwinger--Dyson equation (\ref{SDequation_mcw_p}) in the Fourier space now takes the form (it is enough to consider $x>0$):
\begin{equation}
\sqrt{\bar{M}^2_0 - \partial^2} \, \mathcal{W}_s(x) \, = \,
\frac{g^2}{8\pi} \frac{e^{-x}}{x} \mathcal{W}_s(x) \; .
\label{SDequation_mcw_x}
\end{equation}

Using this equation it is convenient to examine the $\bar{p} \to \infty$ asymptotics. The left-hand side (\ref{SDequation_mcw_p}) has a simple
form $\sqrt{\bar{M}^2_0+\bar{p}^2} \to |\bar{p}|$. The right-hand side of the Schwinger--Dyson equation has a simple form in
the Fourier space (\ref{SDequation_mcw_x}). The limit $\bar{p} \to \infty$ corresponds to the limit $x \to 0$, so the Taylor expansion can be used.

\section{Asymptotics of solution at high momentum.}

\subsection{Contribution from low momentum.\label{section_subsection_low_momentum}}

Let $a$ be a point such that at $\bar{p} > a$ the solution has the asymptotic behavior. If we consider $\bar{p} \gg a$, the contribution from
the right-hand side (\ref{SDequation_final}) from a low $\bar{q}$ is
$$
\frac{g^2}{(4\pi)^2}\frac{1}{\bar{p}} \int\limits_{0}^{a} d\bar{q} \frac{\bar{q} \bar{M}(\bar{q})}{\sqrt{\bar{M}^2(\bar{q})+\bar{q}^2}}
\ln\biggl(1+\frac{4\bar{p}\bar{q}}{1+(\bar{p}-\bar{q})^2}\biggr) \;\simeq\;
\frac{g^2}{4\pi^2} \int\limits_{0}^{a} d\bar{q} \, W(\bar{q}) \bar{q} \cdot \frac{1}{\bar{p}^2} \; .
$$
Consequently, (remind that from (\ref{condition_w_1}) and (\ref{condition_mcw_1}), as $\bar{q} \to +\infty$ all three functions have the same asymptotic
behavior: $\bar{M}(\bar{q}) \simeq W(\bar{q}) \simeq \mathcal{W}(\bar{q})$) the asymptotics of $W(\bar{p})$ cannot be less than $\frac{1}{\bar{p}^2}$:
\begin{equation}
\lim_{\bar{p} \to +\infty} \frac{1}{\bar{p}^2 \, W(\bar{p})} \:<\: +\infty \; .
\label{condition_asym_big}
\end{equation}
Hence, it follows that $W(\bar{p})$ cannot decrease exponentially.

\subsection{Power asymptotics.}

One can examine the following ansatz as $\bar{p} \to \infty$:
\begin{equation}
\mathcal{W}(\bar{p}) \,=\, C \: \mathrm{Sign}(\bar{p}) \, \frac{1}{|\bar{p}|^\beta}
\label{ansatz_power_p}
\end{equation}
where $C$ is some constant.

$\beta$ should be real, otherwise the demand $M(p) \geqslant 0$ is violated. From the requirement of
the convergence of the integral $\int d\bar{q} \, \bar{M}(\bar{q})$ on the upper
limit (see Section \ref{section_analytical_restrictions}) there follows $\beta > 1$. It follows from (\ref{condition_asym_big})
that $\beta \leqslant 2$.

The Fourier-sine transform (\ref{transform_sin-Fourier}) of
the function (\ref{ansatz_power_p}) can easily be calculated, for $0 < \beta < 2$:
$$
\mathcal{W}_s(x) \:=\: C \sqrt{\frac{2}{\pi}} \, \cos\Bigl(\frac{\beta\pi}{2}\Bigr) \: \Gamma (1 \!-\! \beta) \, \frac{1}{x^{1-\beta}} \; .
$$

Substituting this and (\ref{ansatz_power_p}) in (\ref{SDequation_mcw_x}) and (\ref{SDequation_mcw_p}) we get that power asymptotics
for $1 < \beta < 2$ is self-consistent if:
\begin{equation}
\frac{1}{g^2} = \frac{1}{8\pi} \frac{\cot\big(\frac{\beta\pi}{2}\big)}{(1-\beta)} \; .
\label{condition_power_wrong}
\end{equation}
Unfortunately, to obey this formula, one needs $g^2 \leqslant 16$, which is in contradiction with the results of Section \ref{section_analytical_restrictions}.

The value $\beta = 2$ may easily be examined and it does not suit too (see Subsection \ref{section_subsection_certain_cases}).

Combining all together, we have that power asymptotics (\ref{ansatz_power_p}) is not valid for all $\beta$.

\subsection{Log-power asymptotic.}

As in the previous subsection we can test a log-power asymptotics as $\bar{p} \to +\infty$:
\begin{equation}
\mathcal{W}(\bar{p}) \,\simeq\, C \, \frac{(\ln \bar{p})^\gamma}{\bar{p}^\beta} \; .
\label{ansatz_logpower_p}
\end{equation}

From the demand $M(p) \geqslant 0$ it follows that $\beta \in \mathbb{R}$ and $\gamma \in \mathbb{R}$. From the requirement of
the convergence of the integral $\int d\bar{q} \, \bar{M}(\bar{q})$ on the upper
limit (see Section \ref{section_analytical_restrictions}) there follows $\beta > 1, \: \gamma \in \mathbb{R}$ or $\beta = 1, \: \gamma < -1$. To
obey (\ref{condition_asym_big}), we need $\beta < 2, \: \gamma \in \mathbb{R}$ or $\beta = 2, \: \gamma \geqslant 0$.

The Fourier-sine transform (\ref{transform_sin-Fourier}) of some function with asymptotics (\ref{ansatz_logpower_p}) can be expressed in
terms of elementary functions or relatively simple special functions only in a small number of cases of $\gamma$. Fortunately, we do not
need the whole $\mathcal{W}_s(x)$, for our purposes just the asymptotic as $x \to 0$ is sufficient, and it can be calculated
for $0 < \beta < 1$ and $1 < \beta < 2$ and $\gamma \in \mathbb{R}$ (see (\ref{formula_finf_logpower}) in the Appendix):
$$
\mathcal{W}_s(x) \:\simeq\: C \sqrt{\frac{2}{\pi}} \, \cos\Bigl(\frac{\beta\pi}{2}\Bigr) \: \Gamma (1 \!-\! \beta) \: \frac{1}{x^{1-\beta}}\Bigl(\ln\frac{1}{x}\Bigr)^\gamma \; .
$$
This leads to the same constraint (\ref{condition_power_wrong}), so the case $1 < \beta < 2, \: \gamma \in \mathbb{R}$ can not be.

The cases $\beta = 1, \: \gamma < -1$ and $\beta = 2, \: \gamma \geqslant 0$ can easily be considered
directly (see Subsection \ref{section_subsection_certain_cases}), and they also do not suit.

Combining the aforesaid we get that log-power asymptotics (\ref{ansatz_logpower_p}) is not valid for all $\beta$ and $\gamma$.

\subsection{Integral power -- Log asymptotics.\label{section_subsection_certain_cases}}
Consider asymptotic (\ref{ansatz_logpower_p}) in the cases $\beta = 1, \: \gamma < -1$ and $\beta = 2, \: \gamma \geqslant 0$.

Also let $a$ be a point such that for $\bar{p} > a$ the solution $\mathcal{W}(\bar{p})$ has the asymptotic behavior, and we can take $a > 1$. For such $\bar{p}$ the
integral in the right-hand side (\ref{SDequation_final}) can be decomposed into the sum:
$$
\int\limits_{0}^{+\infty} d\bar{q} \; W(\bar{q}) \ln\biggl(\frac{1+(\bar{p}+\bar{q})^2}{1+(\bar{p}-\bar{q})^2}\biggr) \;\simeq\;
\underbrace{\int\limits_{0}^{a} d\bar{q} \; W(\bar{q}) \ln\biggl(1+\frac{4\bar{p}\bar{q}}{1+(\bar{p}-\bar{q})^2}\biggr)}_{{\displaystyle \equiv I_0(\bar{p})}} \:+\:
\frac{C}{\bar{p}^{(\beta-1)}} \int\limits_{\frac{a}{\bar{p}}}^{+\infty} dy \; \frac{\big(\ln (\bar{p}y) \big)^\gamma}{y^\beta}
\ln\biggl(\frac{1+\bar{p}^2(1+y)^2}{1+\bar{p}^2(1-y)^2}\biggr)
$$
where we introduced a new variable $y$ by the formula $\bar{q} \equiv \bar{p} y$. The asymptotics of $I_0(\bar{p})$ was considered in
subsection \ref{section_subsection_low_momentum} and it is proportional to $\frac{1}{\bar{p}}$.

Let us choose $y_1$ and $y_2$ so that $0 < y_1 \ll 1$ and $1 \ll y_2$. Then
the second integral in the right-hand side can be rewritten in the form of the sum: $I_1(\bar{p}) + I_2(\bar{p}) + I_3(\bar{p})$, where
\begin{eqnarray*}
I_1(\bar{p}) &\equiv& \frac{C}{\bar{p}^{(\beta-1)}} \int\limits_{\frac{a}{\bar{p}}}^{y_1} dy \; \frac{\big(\ln (\bar{p}y) \big)^\gamma}{y^\beta} \,
\ln\biggl(1 \,+\, \frac{4y}{\frac{1}{\bar{p}^2} + (1-y)^2}\biggr) \\ [\smallskipamount]
I_2(\bar{p}) &\equiv& \frac{C}{\bar{p}^{(\beta-1)}} \int\limits_{y_1}^{y_2} dy \; \frac{\big(\ln(\bar{p}) + \ln(y) \big)^\gamma}{y^\beta} \,
\ln\biggl(\frac{1+\bar{p}^2(1+y)^2}{1+\bar{p}^2(1-y)^2}\biggr) \\ [\smallskipamount]
I_3(\bar{p}) &\equiv& \frac{C}{\bar{p}^{(\beta-1)}} \int\limits_{y_2}^{+\infty} dy \; \frac{\big(\ln (\bar{p}y) \big)^\gamma}{y^\beta} \,
\ln\biggl(1 \,+\, \frac{4y}{\frac{1}{\bar{p}^2} + (1-y)^2}\biggr) \; .
\end{eqnarray*}
Further, we work only with such $\bar{p}$ that $\frac{a}{\bar{p}} \leqslant y_1, \; \frac{1}{\bar{p}} \ll y_1, \; y_2 \ll \bar{p}$. For this $\bar{p}$ the
integrands are simplified
\begin{eqnarray*}
I_1(\bar{p}) &\simeq& \frac{4C}{\bar{p}^{(\beta-1)}} \int\limits_{\frac{a}{\bar{p}}}^{y_1} dy \; \frac{\big(\ln (\bar{p}y) \big)^\gamma}{y^{(\beta-1)}} \\ [\smallskipamount]
I_2(\bar{p}) &\simeq& \frac{C \big(\ln(\bar{p})\big)^\gamma}{\bar{p}^{(\beta-1)}} \int\limits_{y_1}^{y_2} dy \; \frac{1}{y^\beta} \,
\ln\biggl(\frac{1+\bar{p}^2(1+y)^2}{1+\bar{p}^2(1-y)^2}\biggr) \\ [\smallskipamount]
I_3(\bar{p}) &\simeq& \frac{4C}{\bar{p}^{(\beta-1)}} \int\limits_{y_2}^{+\infty} dy \; \frac{\big(\ln (\bar{p}y) \big)^\gamma}{y^{(\beta+1)}} \; .
\end{eqnarray*}
This integrals can now be calculated directly.

For $\beta = 1, \: \gamma < -1$ this leads to:
\begin{align*}
I_1(\bar{p}) &\simeq 4C \, y_1 (\ln \bar{p})^\gamma &
I_2(\bar{p}) &\simeq C \Big( \pi^2 - 4y_1 - \frac{4}{y_2} \Big) (\ln \bar{p})^\gamma &
I_3(\bar{p}) &\simeq \frac{4C}{y_2} (\ln \bar{p})^\gamma \; .
\end{align*}
Combining all together the right-hand side of (\ref{SDequation_final}) equals: $\displaystyle C \frac{g^2}{16} \frac{(\ln \bar{p})^\gamma}{\bar{p}}$, which
can be consistent with the left-hand side of (\ref{SDequation_final}) only if $g^2 = 16$, but this value is forbidden by the arguments
of Section \ref{section_analytical_restrictions}. Thus, the case $\beta = 1, \: \gamma < -1$ does not suit.

For $\beta = 2, \: \gamma \geqslant 0$ the integration leads to
\begin{align*}
I_1(\bar{p}) &\,\simeq\, \frac{4C}{1+\gamma} \frac{(\ln \bar{p})^{\gamma+1}}{\bar{p}} \,+\, 4C \ln(y_1) \frac{(\ln \bar{p})^\gamma}{\bar{p}} &
I_2(\bar{p}) &\simeq C \Big( 4 - 4\ln(y_1) - \frac{2}{{y_2}^2} \Big) \frac{(\ln \bar{p})^\gamma}{\bar{p}} &
I_3(\bar{p}) &\simeq C \frac{2}{{y_2}^2} \frac{(\ln \bar{p})^\gamma}{\bar{p}} \; .
\end{align*}
From the aforesaid
\begin{equation}
I_0(\bar{p}) + I_1(\bar{p}) + I_2(\bar{p}) + I_3(\bar{p}) \;\simeq\;
\frac{4C}{1+\gamma} \frac{(\ln \bar{p})^{\gamma+1}}{\bar{p}} \,+\, 4C \frac{(\ln \bar{p})^\gamma}{\bar{p}} \,+\,
\frac{4}{\bar{p}} \int\limits_{0}^{a} d\bar{q} \; W(\bar{q}) \bar{q} \; .
\label{condition_log_power_asympt}
\end{equation}
We can see that the right- and left-hand sides of (\ref{SDequation_final}) here are not self-consistent. So the case $\beta = 2, \: \gamma \geqslant 0$ is not valid either.

The $I_0(\bar{p})$ always gives the contribution to asymptotics proportional to $\frac{1}{\bar{p}^2}$. After substitution this asymptotics into the right-side
of (\ref{SDequation_final}), according to (\ref{condition_log_power_asympt}), this should lead to a contribution proportional to $\frac{\ln \bar{p}}{\bar{p}^2}$; after
substitution the last one we should get $\frac{(\ln \bar{p})^2}{\bar{p}}$, and so on. Consequently, we can conclude that condition (\ref{condition_asym_big}) can be generalized to:
$$
\lim_{\bar{p} \to +\infty} \frac{(\ln \bar{p})^\gamma}{\bar{p}^2 \, W(\bar{p})} \:<\: +\infty
$$
where $\gamma \in \mathbb{R}$.

Furthermore, the form of (\ref{condition_log_power_asympt}) suggests that the solution should be searched in the form of a series in powers of the logarithm.

\subsection{Series asymptotics.}

We can suppose that asymptotics of the solution of equation (\ref{SDequation_final}) as $\bar{p} \to +\infty$ has the form
\begin{equation}
W(\bar{p}) \,=\, C_0 \frac{1}{\bar{p}^\beta} + C_1 \frac{\ln (\bar{p})}{\bar{p}^\beta} + C_2 \frac{(\ln (\bar{p}))^2}{\bar{p}^\beta} + \dotsb + L(\bar{p}) \; ,
\label{ansatz_series}
\end{equation}
where $1 < \beta < 2$, the function $L(\bar{p})$ decreases faster than $\frac{1}{\bar{p}^2 \ln (\bar{p})}$, and the series does not reduce
to powers of $\bar{p}$ or $\ln (\bar{p})$.

In the left-hand side of (\ref{SDequation_final}), if we neglect $L(\bar{p})$, then with the same accuracy $\bar{M}(\bar{q}) \simeq W(\bar{q})$, this cames
from the formula inverse to (\ref{def_w}).

In the right-hand side (\ref{SDequation_final}), the integral can be expanded
into the sum: $\int\limits_{0}^{\infty} = \int\limits_{0}^{a} + \int\limits_{a}^{\infty}$ , the integral $\int\limits_{0}^{a}$ was considered
in subsection \ref{section_subsection_low_momentum}, the other integral (to the accuracy of $L(\bar{p})$, which can give not more than $\frac{1}{\bar{p}}$ asymptotic) is
$$
\int\limits_{a}^{\infty} d\bar{q} \; W(\bar{q}) \ln\biggl(\frac{1+(\bar{p}+\bar{q})^2}{1+(\bar{p}-\bar{q})^2}\biggr) \;\simeq\;
\biggl( C_0 \:+\: C_1 \Bigl( -\frac{\partial}{\partial\beta} \Bigr) \,+\: C_2 \Bigl( -\frac{\partial}{\partial\beta} \Bigr)^2 +\: \dotsb \biggr)
\int\limits_{a}^{\infty} d\bar{q} \; \frac{1}{\bar{q}^\beta} \ln\biggl(\frac{1+(\bar{p}+\bar{q})^2}{1+(\bar{p}-\bar{q})^2}\biggr) \; .
$$
The right-hand side integral can also be expanded into the
sum: $\int\limits_{a}^{\infty} = \int\limits_{0}^{\infty} - \int\limits_{0}^{a}$. The $\int\limits_{0}^{a}$ gives the asymptotics $\frac{1}{\bar{p}}$ for
the same reasons as in the low momentum case (see subsection \ref{section_subsection_low_momentum}). The integral $\int\limits_{0}^{\infty}$ can
easily be evaluated by parts
\begin{multline*}
\int\limits_{0}^{\infty} d\bar{q} \; \frac{1}{-\beta \!+\! 1} \Bigl( \frac{\partial}{\partial \bar{q}} \bar{q}^{\, -\beta+1} \Bigr)
\ln\biggl(\frac{1+(\bar{p}+\bar{q})^2}{1+(\bar{p}-\bar{q})^2}\biggr) \:=\;
\frac{2}{\beta - 1} \int\limits_{0}^{\infty} d\bar{q} \; \frac{1}{\bar{q}^{\beta-1}}
\biggl( \frac{\bar{p}+\bar{q}}{1+(\bar{p}+\bar{q})^2} + \frac{\bar{p}-\bar{q}}{1+(\bar{p}-\bar{q})^2} \biggr) \:= \\
\shoveright{ =\; \frac{2 \pi}{\beta - 1} \frac{(1 + \bar{p}^2)^{-\frac{\beta}{2}}}{\sin (\beta\pi)} \biggl( (\bar{p}^2 - 1) \sin\Bigl( \beta \,\mathrm{arccot}(\bar{p}) \Bigr)
\:-\: 2\bar{p} \cos\Bigl( \beta \,\mathrm{arccot}(\bar{p}) \Bigr) \:-\: \bar{p}^2 \sin\Bigl( \beta\pi - (-2\!+\!\beta)\,\mathrm{arccot}(\bar{p}) \Bigr) \:+} \\
+\: 2\bar{p} \cos\Bigl( \frac{\beta\pi}{2} \Bigr) \biggl( \cos\Bigl( \beta \arctan(\bar{p}) \Bigr) \,+\, \bar{p} \sin\Bigl( \beta \arctan(\bar{p}) \Bigr) \biggr)
\:+\: \sin\Bigl( {\textstyle \frac{\beta\pi}{2}} + (-2\!+\!\beta)\arctan(\bar{p}) \Bigr) \biggr) \; .
\end{multline*}
As $\bar{p} \to +\infty$ this leads to:
$$
\int\limits_{0}^{\infty} d\bar{q} \; \frac{1}{\bar{q}^\beta} \ln\biggl(\frac{1+(\bar{p}+\bar{q})^2}{1+(\bar{p}-\bar{q})^2}\biggr) \:=\;
2\pi \frac{\tan \bigl( \frac{(\beta-1)\pi}{2} \bigr)}{\beta - 1} \frac{1}{\bar{p}^{\beta-1}} \,+\,
h(\bar{p},\beta) \frac{1}{\bar{p}^\beta}
$$
where $\lim\limits_{\bar{p} \to +\infty} h(\bar{p},\beta) < +\infty$, so this term can be neglected.

Thus, from various sources the term $A \frac{1}{\bar{p}^2}$ appears in the right-hand side of the equation. We rewrite this term in the form:
$$
A \frac{1}{\bar{p}^2} \;=\; A \frac{1}{\bar{p}^\beta} \:+\: A (\beta \!-\! 2) \frac{\ln (\bar{p})}{\bar{p}^\beta} \:+\:
A \frac{(\beta \!-\! 2)^2}{2!} \frac{(\ln (\bar{p}))^2}{\bar{p}^\beta} \:+\: \dotsb \; .
$$
The absence of this term is the reason why asymptotics (\ref{ansatz_power_p}) and (\ref{ansatz_logpower_p}) are not valid. This term
appears from a middle values of $\bar{q}$ in the integral in the right-hand side of equation (\ref{SDequation_final}).

Finally, the substitution of asymptotics (\ref{ansatz_series}) in equation (\ref{SDequation_final}) leads to the infinite matrix equation
\begin{multline*}
\frac{(4\pi)^2}{g^2}
\left( \begin{array}{c}
C_0 \phantom{\Bigg|} \\ C_1 \phantom{\Bigg|} \\ C_2 \phantom{\Bigg|} \\ \vdots \phantom{\Bigg|}
\end{array} \right)
\;=\;
\left( \begin{array}{c}
A \phantom{\Bigg|} \\ A (\beta \!-\! 2) \phantom{\Bigg|} \\ A \dfrac{(\beta \!-\! 2)^2}{2!} \phantom{\Bigg|} \\ \vdots \phantom{\Bigg|}
\end{array} \right)
\;+ \\ +\;
\left( \begin{array}{cccc}
\: 2\pi \dfrac{\tan \bigl( \frac{(\beta-1)\pi}{2} \bigr)}{\beta - 1} \phantom{\Bigg|} \:&\:
\biggl( -\dfrac{\partial}{\partial\beta} \biggr) \biggl( 2\pi \dfrac{\tan \bigl( \frac{(\beta-1)\pi}{2} \bigr)}{\beta - 1} \biggr) \:&
\biggl( -\dfrac{\partial}{\partial\beta} \biggr)^2 \biggl( 2\pi \dfrac{\tan \bigl( \frac{(\beta-1)\pi}{2} \bigr)}{\beta - 1} \biggr) &\: \cdots \:\\
0 \phantom{\Bigg|} & \dbinom{1}{1} 2\pi \dfrac{\tan \bigl( \frac{(\beta-1)\pi}{2} \bigr)}{\beta - 1} &
\dbinom{2}{1} \biggl( -\dfrac{\partial}{\partial\beta} \biggr) \biggl( 2\pi \dfrac{\tan \bigl( \frac{(\beta-1)\pi}{2} \bigr)}{\beta - 1} \biggr) & \cdots \\
0 \phantom{\Bigg|} & 0 & \dbinom{2}{2} 2\pi \dfrac{\tan \bigl( \frac{(\beta-1)\pi}{2} \bigr)}{\beta - 1} & \cdots \\
\vdots \phantom{\Bigg|} & \vdots & \vdots & \ddots
\end{array} \right)
\left( \begin{array}{c}
C_0 \phantom{\Bigg|} \\ C_1 \phantom{\Bigg|} \\ C_2 \phantom{\Bigg|} \\ \vdots \phantom{\Bigg|}
\end{array} \right)
\end{multline*}
where $\binom{n}{k} = \frac{n!}{k!(n-k)!}$. Likely, one cannot reduce the infinite matrix and columns to finite ones, because the elements in the rows in the matrix do not
decrease. This is one of the reasons why asymptotics (\ref{ansatz_series}) is permitted while
asymptotics (\ref{ansatz_power_p}) and (\ref{ansatz_logpower_p}) are not. The other reason is that in the evaluation of (\ref{ansatz_series}) we do not neglect the
term $A \frac{1}{\bar{p}^2}$.

\section{Conclusions.}

The natural method of obtaining a dimensional parameter in QCD is suggested by means of normal ordering of fields in the Lagrangian. Within our
model, the dimensional parameter QCD is nothing else but the effective gluon mass.

Based on QCD the effective action of strong interaction (\ref{generating_functional_final}) was constructed.

In the framework of the constructed model, the Schwinger--Dyson equation with the effective gluon mass (\ref{SDequation_final}) is investigated both analytically and
numerically. It is shown that spontaneous chiral symmetry breaking occurs
and a nontrivial dependance of the quark mass on momenta. The critical value of the strong coupling constant (equals to $\alpha_s = 4/{\pi}$), \emph{above} which the
spontaneous breaking occurs, is found in the semiclassical approximation. It is proved
strictly that \emph{below} this critical value, the Schwinger--Dyson equation has only trivial non-negative solution $M(p)=0$.

Although the derivation of the effective action of strong interaction (\ref{generating_functional_final}) from the QCD Lagrangian (\ref{lagrangian_QCD_initial}) is clear
and well-controlled, numerous assumptions are done during the derivation. Thus, the obtained results are qualitative rather than quantitative. Taking into account
the neglected terms could amend the model and make it quite quantitative. For better understanding of the solution of the Schwinger--Dyson equation in
the region of the large coupling constant, it would be better to improve the numerical computation scheme. For instance, the accuracy of the numerical simulations is
likely to be insufficient to obtain the precise value of $M(p)$ at $p=0$.

The developed analytical methods of solving and analyzing as well as created programs
for numerical calculation the Schwinger--Dyson equation can be used not only in the considered specific kernel but for various other kernels.

The Fourier-sine transform of function (\ref{def_f}) with log-power asymptotic was performed (\ref{formula_sintransform_lnpow_all}), and the leading asymptotic was found.

In the papers \cite{ShilinPervushin13_QED,PervushinShilin13_QCD} (see also \cite{PervushinReinhardtEbert79,KKPervushin90_F,KKPervushin89_P,Pervushin12_proc} and
references therein) it is shown how the Bethe--Salpeter equation, which describe the spectrum and wave functions of the bound states, can be derived in the framework of
the Stationary Phase method. To cope with this Bethe--Salpeter equation, one should already have the solution of the
corresponding Schwinger--Dyson equation (\ref{SDequation_definition}) as the ``input function''. The investigation of the Bethe--Salpeter equation is beyond the
scope of this paper and will be done later.
%Moreover in the article \cite{PervushinShilin13_QCD} it was noted that in the limit of zero current quark mass the Schwinger--Dyson equation
%after some change of variables coincide with the corresponding Bethe--Salpeter equation.

\begin{acknowledgments}
The authors thank Alexander Cherny, Andrej Arbuzov and Alexander Dorokhov for fruitful discussions.
\end{acknowledgments}

\appendix*
\section{Fourier-sine transform of a log-power function.}

Consider the function
\begin{equation}
F_a (x, \gamma, \beta) \equiv \int\limits_0^{+\infty} d\bar{p} \; \frac{\bigl( \ln ( a \! + \! \bar{p} ) \bigr)^\gamma}{{(a \! + \! \bar{p})}^\beta} \sin(x\bar{p}) \; ,
\label{def_f}
\end{equation}
where: $a>1$, $\gamma \in \mathbb{R}$, $\beta \in \mathbb{R}$. As: $F_a (-x, \gamma, \beta) = - F_a (x, \gamma, \beta)$, below we will take: $x>0$. Our aim is to find the
asymptotic behavior of $F_a (x, \gamma, \beta)$ at $x \to 0$.

This function has the property
\begin{equation}
F_a (x, \gamma, \beta) = -\frac{d}{d\beta} F_a (x, \gamma\! - \! 1, \beta) \; ,
\label{formula_gamma_from_beta}
\end{equation}
so the practically interesting case is: $ -1 < \gamma \leqslant 0$. Let us introduce the new notaion: $\bar{\gamma} \equiv -\gamma$, where: $0 \leqslant \bar{\gamma} < 1$.

Using the formula
$$
\frac{1}{\bigl( \ln ( a \! + \! \bar{p} ) \bigr)^{\bar{\gamma}}} = \int\limits_0^{+\infty} dt \; \frac{\sin \bigl( \, t \, \ln ( a \! + \! \bar{p} ) \bigr)}{t^{1 - \bar{\gamma}}}
\frac{1}{\sin \bigl( \frac{\pi \bar{\gamma}}{2} \bigr) \: \Gamma(\bar{\gamma})}
$$
and exchanging the order of integrations we get
\begin{equation}
F_a (x, - \bar{\gamma}, \beta) = \frac{1}{\sin \bigl( \frac{\pi \bar{\gamma}}{2} \bigr) \: \Gamma(\bar{\gamma})} \int\limits_0^{+\infty} dt \; \frac{1}{t^{1 - \bar{\gamma}}}
\int\limits_0^{+\infty} d\bar{p} \; \frac{\sin(x\bar{p})}{{(a \! + \! \bar{p})}^\beta} \: \sin \bigl( \, t \, \ln ( a \! + \! \bar{p} ) \bigr) \; .
\label{formula_repres_sinf_in_int}
\end{equation}

One can expand the
function: $\sin \Bigl( \, t \, \ln ( a \! + \! \bar{p} ) \Bigr) = \sin \Bigl( \, t \, \ln \bigl( x ( a \! + \! \bar{p} )\bigr) + \, t \, \ln \dfrac{1}{x} \Bigr)$ into the
series around the point $t \, \ln \dfrac{1}{x}$. Put by definition:
$$
H_a (x,\beta) \equiv \int\limits_0^{+\infty} d\bar{p} \; \frac{\sin(x\bar{p})}{{(a \! + \! \bar{p})}^\beta} \; ,
$$
we arrive at
\begin{multline*}
\int\limits_0^{+\infty} d\bar{p} \; \frac{\sin(x\bar{p})}{{(a \! + \! \bar{p})}^\beta} \: \sin \bigl( \, t \, \ln ( a \! + \! \bar{p} ) \bigr) \; = \\
=\; x^\beta \sin \bigl( \, t \, \ln \dfrac{1}{x} \bigr) \sum_{k=0}^\infty \frac{(-1)^k}{(2k)!} \biggl( \frac{d^{2k}}{d\beta^{2k}} \frac{H_a (x,\beta)}{x^\beta} \biggr) t^{2k} \;-\;
x^\beta \cos \bigl( \,t\, \ln\dfrac{1}{x} \bigr) \sum_{k=0}^\infty \frac{(-1)^k}{(2k\!+\!1)!} \biggl( \frac{d^{2k+1}}{d\beta^{2k+1}} \frac{H_a (x,\beta)}{x^\beta} \biggr) t^{2k+1}\;.
\end{multline*}

Substituting the latter into (\ref{formula_repres_sinf_in_int}) and changing the variable: $\bar{t} \equiv t \, \ln \frac{1}{x}$, we finally have
\begin{multline}
F_a (x, - \bar{\gamma}, \beta) \;=\; \frac{1}{\sin \bigl( \frac{\pi \bar{\gamma}}{2} \bigr) \: \Gamma(\bar{\gamma})} \frac{x^\beta}{\bigl( \ln \frac{1}{x} \bigr)^{\bar{\gamma}}}
\int\limits_0^{+\infty} d\bar{t} \; \Biggl( \sum_{k=0}^\infty \frac{(-1)^k}{(2k)!} \biggl( \frac{d^{2k}}{d\beta^{2k}} \frac{H_a (x,\beta)}{x^\beta} \biggr)
\frac{1}{\bigl( \ln \frac{1}{x} \bigr)^{2k}} \: \bar{t}^{\, 2k - 1 + \bar{\gamma}} \sin \bar{t} \; - \\
- \; \sum_{k=0}^\infty \frac{(-1)^k}{(2k\!+\!1)!} \biggl( \frac{d^{2k+1}}{d\beta^{2k+1}} \frac{H_a (x,\beta)}{x^\beta} \biggr)
\frac{1}{\bigl( \ln \frac{1}{x} \bigr)^{2k+1}} \: \bar{t}^{\, 2k + \bar{\gamma}} \cos \bar{t} \Biggr) \; .
\label{formula_sintransform_lnpow_all}
\end{multline}

Further steps strongly depend on which $\beta$ we want to consider and how many terms of the series we are interested in. In our case, we are interested in only leading
asymptotics and values: $0 < \beta < 1$ or $1 < \beta < 2$. For these $\beta$:
$$
H_a (x,\beta) \simeq \frac{\pi \cos \bigl( \frac{\beta \pi}{2} \bigr)}{\Gamma (\beta) \sin (\beta \pi)} \frac{1}{x^{1-\beta}}
$$
and leading asymptotics comes from the $k=0$ term in the first sums (\ref{formula_sintransform_lnpow_all}). Thus, we get
$$
F_a (x, - \bar{\gamma}, \beta) \simeq
\frac{\pi \cos \bigl( \frac{\beta \pi}{2} \bigr)}{\Gamma (\beta) \sin (\beta \pi)} \frac{x^{\beta - 1}}{\bigl( \ln \frac{1}{x} \bigr)^{\bar{\gamma}}} \; .
$$

Recollecting (\ref{formula_gamma_from_beta}) we finally have
\begin{equation}
F_a (x, \gamma, \beta) \;\simeq\; \Gamma (1 \!-\! \beta) \cos \Bigl( \frac{\beta \pi}{2} \Bigr) \: x^{\beta - 1} \, \Bigl( \ln \frac{1}{x} \Bigr)^\gamma \; ,
\label{formula_finf_logpower}
\end{equation}
where: $x>0$, $0 < \beta < 1$ or $1 < \beta < 2$, $\gamma \in \mathbb{R}$. The coefficient does not depend on $\gamma$, $a>1$ and $a$ is absent in the
right-hand side, as it should be.

\bibliography{SchwingerDyson}

\end{document}